\documentclass[12pt,preprint]{aastex}

\newcommand{\bc}{\begin{center}}
\newcommand{\ec}{\end{center}}
\newcommand{\be}{\begin{enumerate}}
\newcommand{\ee}{\end{enumerate}}
\newcommand{\beq}{\begin{equation}}
\newcommand{\eeq}{\end{equation}}
\newcommand{\beqa}{\begin{eqnarray*}}
\newcommand{\eeqa}{\end{eqnarray*}}
\newcommand{\bi}{\begin{itemize}}
\newcommand{\ei}{\end{itemize}}
\newcommand{\bt}{\begin{tabular}}
\newcommand{\et}{\end{tabular}}
\newcommand{\etal}{{et~al.\/}\ }
\newcommand{\eg}{{e.g.\/}\ }

\newcommand{\hr}{$^{\rm h}$}
\newcommand{\mn}{$^{\rm m}$}
\newcommand{\st}{$^{\rm s}$}

\newcommand{\ha}{$\rm H{\alpha} $ }
\newcommand{\mh}{ \cal  M_{\rm HI}  }
\newcommand{\mo}{ \cal  M_{\odot}  }
\newcommand{\kms}{$\rm km \ s^{-1}$}

\newcommand{\cld}{the Vela Cloud}
\newcommand{\clds}{the Vela Cloud\ }
\newcommand{\Cld}{The Vela Cloud}
\newcommand{\Clds}{The Vela Cloud\ }

\begin{document}
 
\slugcomment{To be submitted to A.J.}
\shorttitle{The Vela Cloud}
\shortauthors{English \etal}

\title{\Cld: A Giant HI Anomaly in the NGC~3256 Group}

\author{Jayanne English}
\affil{University of Manitoba}
\authoraddr{Department of Physics and Astronomy, University of
Manitoba, Winnipeg, Manitoba, Canada R3T 2M8}

\author{B. Koribalski}
\affil{Australia Telescope National Facility, CSIRO}
\authoraddr{PO Box 76, 
Epping, NSW 1710,
Australia}

\author{J. Bland-Hawthorn}
\affil{Sydney Institute for Astronomy}
\authoraddr{School of Physics A28, University of Sydney, NSW 2006, Australia}

\author{K. C. Freeman}
\affil{Research School of Astronomy and Astrophysics, The Australian
National University}
\authoraddr{Mount Stromlo Observatory, Private Bag, Weston Creek P.O.,
  A. C. T. ~2611, Australia}

\author{Claudia F. McCain}
\affil{Research School of Astronomy and Astrophysics, The Australian
National University}
\authoraddr{Molenstraat 24, 3764 TG Soest, The Netherlands}

\begin{abstract}

  We present Australia Telescope Compact Array (ATCA) observations of a
  galaxy-sized intergalactic HI cloud (``\cld'') in the NGC~3256 galaxy group.
  The group contains the prominent merging galaxy NGC~3256, which is surrounded
  by a number of HI fragments, the tidally disturbed galaxy NGC~3263, and
  several other peculiar galaxies.  \Cld, with an HI mass of 3-5 $\times \ 10^9
  \mo$, resides southeast of NGC~3256 and west of NGC~3263, within an area of
  9\arcmin\ $\times$ 16\arcmin\ (100 kpc $\times$ 175 kpc for an adopted
  distance of 38 Mpc).  In our ATCA data \clds appears as 3 diffuse components
  and contains 4 density enhancements.  \Cld's properties, together with its
  group environment, suggest that it has a tidal origin.  Each density
  enhancement contains $\rm \sim 10^{8} \mo$ of HI gas which is sufficient
  material for the formation of globular cluster progenitors.  However, if we
  represent the enhancements as Bonnor-Ebert spheres, then the pressure of the
  surrounding HI would need to increase by at least a factor of 6 in order to
  cause the collapse of an enhancement.  Thus we do not expect them to form
  massive bound stellar systems like super star clusters or tidal dwarf
  galaxies.  Since the HI density enhancements have some properties in common
  with 
  High Velocity Clouds, we explore whether they may evolve to be identified with
  these starless clouds instead.

\end{abstract}

\begin{figure}
\epsscale{.7}
\plotone{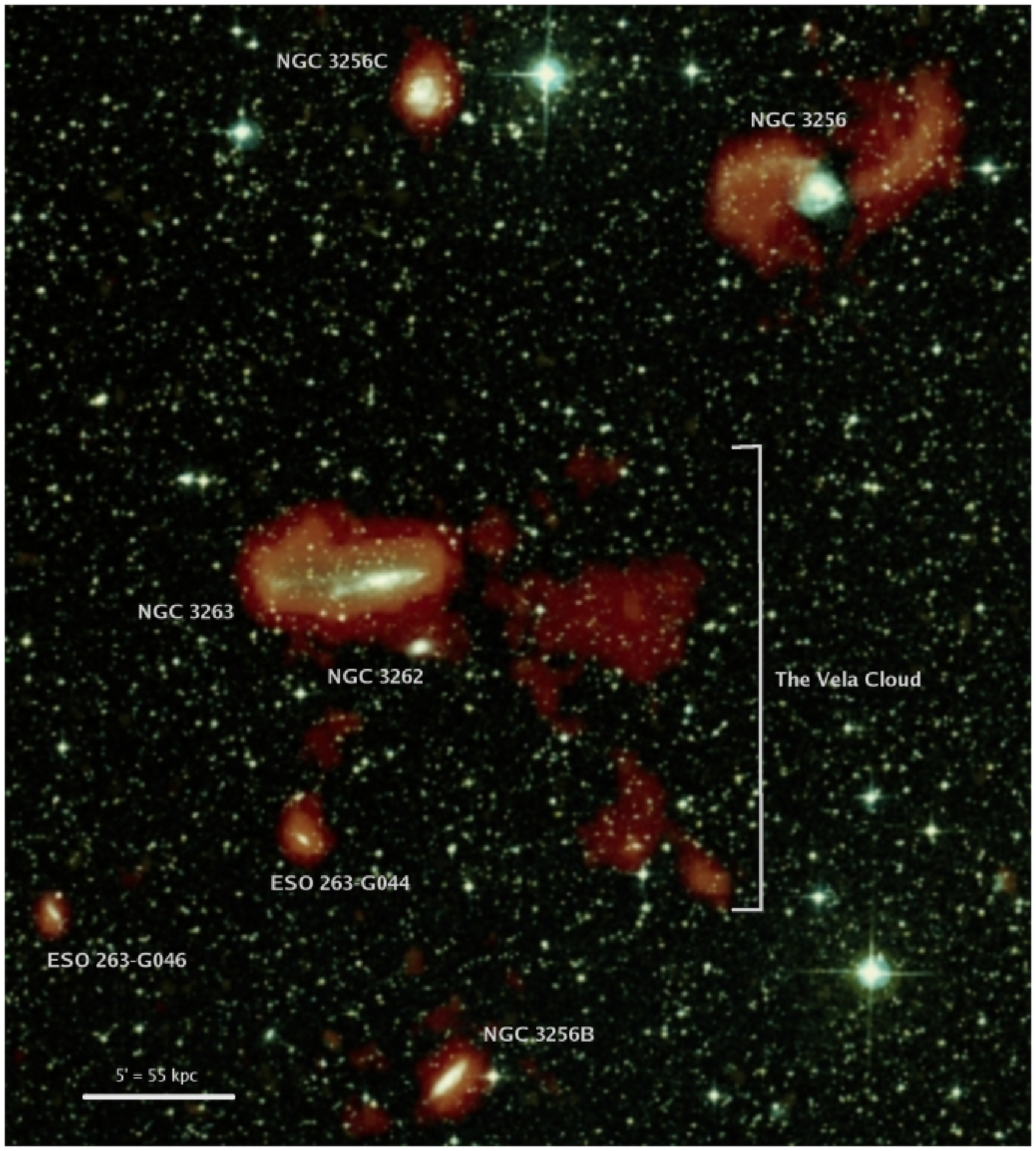}
\epsscale{1}
\figcaption[f1.eps]{HI distribution in the NGC~3256
Group of galaxies.  The HI distribution shown is the  combination of
2 different ATCA moment maps,  both of which are integrated 
from 2621 to 3337 \kms.  One map was generated from a cube with a
naturally weighted synthesized beam
of 83\arcsec\ $\times$ 67\arcsec\  while the other map is from a cube with
an approximately uniformly
weighted beam with dimensions 58\arcsec\ $\times$ 67\arcsec. 
The initial greyscale intensity stretches were logarithmic in order to
emphasize fainter emission.  The lower resolution map was assigned a darker
red than the higher resolution map. The maps were then combined using 
the screen algorithm in ``The GIMP'' image editing
package; The GIMP is written by Peter Mattis and Spencer Kimball, 
and released under the GNU General Public License.
We used image editing techniques described in
\cite{rlfek04}. The HI emission maps have been overlaid on
colourized blue and red Digitized Sky Survey
images. North is up and east is to the left. 
In these ATCA data, \clds appears as 3 components to the west 
of NGC~3263,  which is the edge-on galaxy near the center of the image.
The cloud is plotted in more detail, with coordinates, 
in Fig.~\ref{zoomcloud}.\label{HIoverlay}}
\end{figure}

{

\section{Introduction}
\label{intro}

\subsection{Intergalactic clouds}
\label{introclouds}

The detection of intergalactic clouds, such as the one studied in this paper,
can illuminate the progress of galaxy evolution.  For example, one expects to
find primordial HI building blocks of hierarchical galaxy construction unless
galaxy formation was extremely efficient. However these are rarer than predicted
by the current cosmological theories which build galaxies via the hierarchical
merging of proto-galactic clouds embedded in dark-matter halos
(\eg\citealt{1999ApJ...522...82Klypin}).  Studies of groups of galaxies reveal
very few metal-poor and star-poor clouds (\eg \citealt{2001MNRAS.325.1142Zwaan}
and papers referenced therein; \citealt{2004ApJ...610L..17Pisano}). Other HI
detections include bona fide dwarf galaxies
(\eg\citealt{2000AJ....120.3027Cote}); HI envelopes enclosing superwinds (\eg
\citealt{2005MNRAS.358.1453Ott}); and outflows from disks due to supernovae or
buoyantly rising bubbles (\eg\cite{2003ApJ...585..268Irwcha} and papers
mentioned therein) or plumes potentially due to ram pressure
(\eg\citealt{2005AA...437L..19OosVG}). A strongly accepted explanation for many
HI clouds is that they are tidal extensions from galaxies and/or tidal dwarf
galaxies (\eg \citealt{2004A&A...427..803DucBour};
\citealt{2001AJ....122.2969Hibbard}; \citealt{2001ASPC..240..657Hib};
\citealt{2003MNRAS.339.1203Kor}; \citealt{2004MNRAS.348.1255Kor};
\citealt{IAUS..217...34Kor}).  Tidal origins are also proposed for 
HI companions to HII galaxies (Taylor et
al. (1993, 1995,
1996)\nocite{1993AJ....105..128Taylor}\nocite{1995ApJS...99..427Taybrin}\nocite{1996ApJS..102..189Taylor}, 
the compact HI companion to FCC~35 \citep{1998AJ....115.2345Putman}, and
the extended companion to NGC~2442 called HIPASS J0731-69
\citep{2001ApJ...555..232Ryder}.  Indeed the Leo Ring
\citep{1983ApJ...273L...1Schneider}
is associated with a group of galaxies and is likely to
have a tidal origin \citep{2005MNRAS.357L..21BekKor}.

A similar debate also exists about the origins of the Galactic High Velocity
Clouds (HVCs).  Since stars in HVC have not yet been detected, their distances
from the Milky Way are difficult to determine.  Recent studies indicate that
they are within tens of kpc (Bland-Hawthorn et al 1998; Putman et al 2003$a$;
\citealt{2008ApJ...672..298Wakker}).  HVCs could be
nearby tidal debris similar to fragments of the Magellanic Stream
(\eg\citealt{2008MNRAS.388L..29WestKor}, \citealt{2003ApJ...586..170PutSS}) or
material cooling and infalling from a primordial halo
(\eg\citealt{2008ApJ...674..227PeekPutSom} although see
Binney, Nipoti \& Fraternali 2009). Another alternative is that they are material ejected from the disk of
our Galaxy to a height of a few kpc which subsequently rains back down onto the
plane as a Galactic fountain (\eg\citealt{1989ApJ...345..372NorIke};
\citealt{2005ApJS..157..251GangSemb}).

In order to clarify the origins of HVCs, there are numerous studies of
extra-planar HI associated with individual galaxies.
Since the detected HI clouds
are in the vicinity of the host galaxies the origins of these 
clouds are interpreted as either a galactic fountain (e.g.~\citealt{2008MNRAS.386..935FratBin}) or tidal debris
(e.g.~\citealt{2007NewAR..51..108WestThilk}) or possibly a condensation from a
primordial halo (e.g.~\citealt{2008ApJ...676..991RandBen}).

A survey for extra-galactic HVCs in groups of galaxies by
\nocite{2004ApJ...610L..17Pisano}\nocite{2007ApJ...662..959Pisano} Pisano et
al.~(\eg 2004, 2007, and papers mentioned therein) have detected optical
counterparts to HI clouds, suggesting they are dwarf galaxies.  Also the
proximity of the dwarfs to individual galaxies within each group seems to argue
against in-fall from outside a group's radius, making a cosmological origin
unlikely.  That is, we would expect primordial cold clouds to fall towards
the group's centre of mass rather than predominantly
orbit individual galaxies.
The on-going merger NGC~3256 
also displays HI fragments in its vicinity
\citep{2003AJ....125.1134EngNor}. However since they appear to be starless, they
may be analogous to HVCs.

\subsection{\Clds in the NGC~3256 Group}
\label{cloud}

The NGC~3256 Group and NGC~3263 Group are difficult to distinguish from each
other spatially and the systemic velocities of their eponymous galaxies differ
by less than 200 \kms. Additionally two different group-finding algorithms
applied to the same dataset do not find the same galaxies in each group
({\citealt{1992A&AS...93..211Fouque}; \citealt{1993A&AS..100...47Garcia}).  Many
researchers therefore consider at least 15 galaxies to be members of a single
large group spread over roughly a few degrees
(eg.~\citealt{2000AJ....120..645Lipari}; \citealt{1991ApJ...383..467Mould}). 
We adopt this latter perspective for the remainder of the paper and refer
to the association of galaxies as the NGC~3256 Group.

The most spectacular HI feature in the NGC~3256 Group is a galaxy-sized
intergalactic HI cloud \citep{eng94}, which we will refer to as ``\cld''.  The
structure, as projected on the sky, is not clearly associated with an individual
galaxy but appears to be part of the group. A tidal origin for the cloud is
suggested by the fact that most of the galaxies in this region
are tidally disturbed. This fact can be seen from those members of the group that are
present in the field of view of our  Australia Telescope Compact Array (ATCA)
HI data  (Fig.~\ref{HIoverlay}).   
The cloud's
potential evolution is also interesting, 
and this paper explores whether parts of the cloud could be identified with HVCs
at some point in their evolution.

\Clds was initially detected in an ATCA pointing towards NGC~3256
\citep{eng94,2003AJ....125.1134EngNor} and further studied with 3
pointings of the Parkes Telescope \citep{eng94}.  In this paper we
present additional ATCA data with a pointing centre towards NGC~3263, which
confirms \cld's existence (Fig.~\ref{HIoverlay}).  \Clds
appears to be a cohesive structure, with 3 distinct diffuse gas
sub-components, which we have labelled A, B, and C; see the schematic
in Fig.~\ref{schematic}.  
Four HI density enhancements are seen and 
an aim of this paper is to consider whether the internal pressure
within these enhancements satisfies the conditions for initiating
star formation.

Our radio and optical (broad band and Fabry-Perot) observations and reductions
are described in \S~\ref{observations}.  \S~\ref{21measure} only describes
measurement methods and the uncertainties associated with these methods.  The
actual measurements of the cloud's observed characteristics are detailed in
section \S~\ref{results}.
In \S~\ref{derived} we describe derived characteristics (such as
the amount of HI mass in emission), explore the star-forming potential
in \clds (representing the enhancements as Bonnor-Ebert spheres), and consider
the ultraviolet radiation field 
of NGC~3263 in relation to  \cld. 
Then in \S~\ref{compare} the cloud's characteristics are contrasted
with those of tidal debris, and other
apparently isolated clouds, including HVCs.  
In \S~\ref{discussion}, we discuss the likely role of the group's
tidal field on the origin of \clds and speculate on the cloud's
fate, asking whether its enhancements may evolve into HVCs. 
\S~\ref{summary} summarizes our study of this rare galaxy-sized
intergalactic cloud.

\section{Observations}
\label{observations}

\subsection{Radio Observations and Reductions.}
\label{radioobs}

21-cm line data were acquired using the ATCA in 1995 and 1996.  The
target centre was right ascension 
$\alpha$~=~10\hr 29\mn 04\st\ and 
declination $\delta$~=~-44\degr 04\arcmin12\arcsec (J2000),
close to NGC~3263.  At 1406 MHz, with a bandwidth of 8 MHz, the
observations spanned that galaxy's systemic and rotation velocities
and included several other galaxies in the NGC~3256 Group.  Array
configurations 375, 750A, 1.5B, 6C were used to acquire 12.5, 10.2,
11.6, 10.7 hours, respectively. 
The longest baseline corresponds to a spatial resolution of 7 arcsec
while 
the primary beam is 33 arcmin.  These 4 UV
datasets were flagged and calibrated, using AIPS and the calibrators 
PKS 1934-638 (primary) and PKS 0823-500 (secondary), in the usual way.
The continuum emission was fit using the line-free emission on either
side of the HI signal and subtracted in the UV plane.  After
converting their frequency axes to optically defined heliocentric
velocities, the datasets from the 4 array-configurations were combined.

For the subsequent Fourier transformation, deconvolution, and analysis we used
the {\sc Miriad} software package. We base our data analysis on a cube with the
parameters listed in Table~\ref{atcalog} and on subcubes which span smaller
velocity ranges (see \S~\ref{21measure}).  The characteristics of the cubes used
to produce Fig.~\ref{HIoverlay} and Fig.~\ref{velfldoverlay} are listed in those
figure captions. Although radio continuum observations were also made at 1380
MHz, the intergalactic cloud was not detected.

Single-dish telescope data were acquired on Dec. 19 and 20, 1993 using
the single-feed 20-cm
receiver at the 64-m Parkes Telescope, a bandwidth of 32 MHz, and
2 polarizations of 1024 channels each.
The effective channel width is 8.2 \kms.  Observations were acquired for 3
different target positions on the sky.  For each position 2 scans were acquired
on-target along with 2 reference scans off-target. Each scan was 10 minutes. The
off-target scans were subtracted from the on-target data and also used for
normalization of the spectra; the r.m.s.~is 0.013 Jy per channel. We present one
of the target positions (${\alpha}$~=~10\hr 28\mn 27\fs7, ${\delta}$~=~-44\degr
09\arcmin 25\arcsec (J2000)) in this paper.  While this pointing is west of
NGC~3263, the 15\arcmin\ beam width encompasses some emission from this galaxy.
Details, and profiles of the other pointings, are provided in \cite{eng94}.

\subsection{Optical Observations and Reductions}
\label{opticalobs}

On the night of 27-28 Feb 1995 we observed \clds with the TAURUS
imaging Fabry-Perot interferometer mounted at the f/8 Cassegrain focus
of the AAT 3.9m.  We attempted to detect the HI cloud in H$\alpha$ and
[NII]6583\AA\ emission using the Fabry-Perot `staring' method which
reaches the faintest possible levels in diffuse line emission
\citep{1994ApJ...437L..95Bland94}. The conditions were mostly photometric
and dark, with the moon rising at the end of the night.  We used the
University of Maryland etalon with a 41 $\mu$m gap spacing and a
coating finesse of 50 at H$\alpha$. The blocking filter, which was
centred at 6641\AA\ with a 17\AA\ bandpass, had a peak transmission of
73\%. The measured free spectral range was 54.3\AA\ at H$\alpha$ such
that the blocking filter isolated only one order.  We employed a TEK
1024$^2$ CCD with 0.594" pixels and a read noise of 2.3 electrons in
XTRASLOW mode.  The technique results in observations acquired in an 
annular region which was centred at 
$\alpha$~=~10\hr 28\mn 30\fs2, 
$\delta$~=~-44\degr 08\arcmin  51\arcsec (J2000).
The integration times are listed in Table~\ref{aaocatalog}.

The images were bias subtracted, flatfielded and corrected
for filter transmission profile using twilight and whitelight frames
respectively.  Stars within the field were located with {\tt daophot}
and subsequently subtracted.  The four data frames were combined and
co-added using minmax clipping under {\tt imcombine} in IRAF; this
procedure successfully removed all cosmic rays and point-like
artefacts. The data were azimuthally binned and corrected for the
angular dispersion as discussed in \citet{1998MNRAS.299..611BlandVeil}.  The
final spectrum is dominated by two OH sky-lines at 6627.6\AA\ and 
6634.2\AA. No H$\alpha$ or [NII] emission can be detected at the
level of EM(H$\alpha$) $=$ 60 mR and EM([NII]) $=$ 30 mR (3$\sigma$).
Note that at 10$^4$K, EM(H$\alpha$) = 30 mR is equivalent to a surface
brightness of $2.0 \times 10^{-19}$ erg cm$^{-2}$ s$^{-1}$
arcsec$^{-2}$.

Using the Double-beam Spectrograph at the Siding Spring Observatory
2.3m telescope, a spectrum (\S~\ref{resultsenhance})
was acquired at $\alpha$~=~10\hr 28\mn 25\fs77, 
$\delta$~=~-44\degr 16\arcmin 17\farcs5 (J2000), at the
position of the HII emission-line galaxy WPV060.  The wavelength range
is 6400-6850 \AA, allowing the acquisition of 5 emission lines:
H$\alpha$, [NII], [SII]. Using the $\sigma$ of the calibration arclamp
lines gives an uncertainty for the systemic velocity of 6 \kms.  The
instrumental profile is 4.8\AA\ which corresponds to a velocity
resolution of $\sim 220$ \kms.  The reductions were carried out in the
usual way using IRAF. For the spectrum and its analysis see \S~\ref{resultsenhance}.

\section{21 cm Emission-line Measurement and Analysis Techniques}
\label{21measure}

As described in \S~\ref{cloud}, \clds appears in our ATCA data as 3 large
diffuse components which we have labelled A, B, and C (Figure~\ref{schematic}).
This current section describes the techniques, including the uncertainties, used
to measure the observed characteristics of these diffuse component clouds
(\S~\ref{21measuremom}) and the enhancements embedded in them
(\S~\ref{obsenhancements}).  The measurements themselves are presented in
section \S~\ref{results}.

\subsection{The Diffuse Emission in Components A, B, and C}
\label{21measuremom}

For each diffuse cloud component sketched in Fig~\ref{schematic}, the
Full-Width-Half-Maximum (FWHM) value 
(\S~\ref{detectcloud}) was determined from velocity profiles
which were generated interactively using the {\sc Karma} visualization
package \citep{1996ASPC..101...80Gooch}. An uncertainty of 10-15\%
includes the difference between measuring smoothed and unsmoothed
profiles and the variation produced by visually selecting slightly
different sized rectangles to encompass the component.

Dimensions and integrated flux densities were measured from integrated intensity
(i.e. zero moment) maps generated using {\tt moment} in {\sc Miriad}. All
measurements were made on data corrected for the primary beam response; however
in the case of integrated intensity maps, the primary beam correction was
applied to each {\tt moment} map.
Using cloud component B, we compared this approach to constructing moment maps
from primary beam corrected cubes.  For these comparison maps, a smoothed mask
cube was applied to a primary beam corrected cube and values above a 3 $\sigma$
in the mask are retained and then a moment map was created from the resultant
blanked cube.  To create the mask cube, the primary beam corrected data were
spatially smoothed using a Gaussian of 3 beam widths and Hanning smoothed in
velocity using 3 channels (see Table~\ref{21measure}). We note that \clds is
clearly evident in this mask cube.
This second approach gave an integrated flux density for component B that was
14\% higher than the value from the primary beam-corrected {\tt moment} map
approach.  Since the primary beam-corrected {\tt
  moment} map approach generates smaller values, which will produce conservative
mass estimates, we list these for the
integrated flux density, $\rm \int S \ dv$ (in Jy $\times$ km s$^{-1}$), in
Table~\ref{combinedtable} and in ~\S~\ref{detectcloud}.
The difference between measuring the flux density using {\tt
  moment} maps (in which weak features may be clipped) and summing
channel-by-channel 
(in which noise may also be summed) can be as large as 30\%, and we use this as
our uncertainty.
The uncertainty in the values of each component's dimensions is about 10\% and
reflects the difficulty in determining a component's boundary. 
 
To calculate the HI mass in emission, {$\rm {\cal M}_{HI}$}, for each diffuse
component we use the integrated flux density. To be consistent with the
analysis of NGC~3256 in \cite{2003AJ....125.1134EngNor}, we use a distance of
37.6 Mpc, adopting H$_o$ of 75 \kms\ Mpc$^{-1}$. (The resultant linear scale is
182 pc per 1\arcsec.)  Combining the uncertainty in the flux density above with
an uncertainty of 10\% for the distance, increases the total uncertainty in the
HI mass to at least 33\%.

The column density of each component, used for comparisons with HI
clouds in other systems and objects within the NGC~3256 group
(\S~\ref{comparediffuse}), is also listed in
Table~\ref{combinedtable}. 
For the polygonal area delineating each cloud component, we calculated the mean
intensity S for the number of channels associated with the velocity range listed
in Table~\ref{combinedtable}.
However in order to calculate column density one must measure the brightness temperature
${\rm T_B}$, rather than S.  Using the Rayleigh-Jeans approximation, \beq {\rm T_B =
  \frac{\lambda^2 \ B }{2k}} \eeq where k is the Boltzmann constant and B the
brightness gives \beq{\rm T_B = 1541 \ \lambda({\rm cm})^2 \ B}\eeq in Kelvin.
The 
value for S, which is in Jy beam$^{-1}$, can be converted to B
 by describing the synthesized beam as a factor of
($\pi/(4ln2)$) times the major and minor beam axes.
Therefore, in terms of the measured parameters, the 
column density  (e.g.~equation $3$ of \citealt{1990ARAA..28..215DickLock})
becomes
\beq {\rm N_{HI}\ =\ }{\rm 1.823 \times 10^{18} \times \Delta v \times
no.~channels \times S \times \frac{1\ beam}{1.1331\ \theta_{major}
\times \theta_{minor}} \times 1541 \times \lambda^2}\eeq 
in atoms
cm$^{-2}$.  The channel width $\rm \Delta v$, in \kms, and the axes of
the synthesized beam $\theta$, in arcsec, are presented in
Table~\ref{atcalog}, and the rest wavelength of the observations,
$\lambda$, is 21 cm.  An uncertainty of $<$ 15\% in the column density
arises mainly from the selection of the number of channels associated
with a given diffuse component.

The value of the peak (or maximum) column density within the total
spatial region of each diffuse cloud component was determined
from unclipped {\tt moment} maps. 
Comparison with the clipped moment maps
results in an uncertainty of about 4\%. 

Note that the same techniques above were used to measure other objects in the
dataset such as NGC~3263 and the apparent bridge between it and ESO
263-G044. For example, a {\tt moment} map, constructed only using absolute
values greater than 3 mJy beam$^{-1}$, is used for the bridge's integrated flux
density while both this map and a channel-by-channel calculation constrain its
possible range of column density values (\S~\ref{comparediffuse}). 

\subsection{Detecting  Enhancement Candidates Within Each Component}
\label{obsenhancements}

There is evidence that Cloud B and the southern part of Cloud C  host intensity
enhancements. As an initial detection criterion, the potential HI enhancements
were required to appear morphologically continuous in adjacent channels of the
data cube. Additionally they have flux density peaks that are at least 6
$\times$ r.m.s.~in the HI cube. 

In order to distinguish enhancements from random
fluctuations in the diffuse components which are their hosts, we measured the mean
flux density of the host 
(3 mJy beam$^{-1}$ in both B and C) and the host's
r.m.s.~(2 mJy beam$^{-1}$ and 3 mJy beam$^{-1}$ for B and C
respectively). The enhancement was considered a robust candidate if its 
flux density peak was more than twice this r.m.s.~above the mean flux density of
their host component; this signal-to-noise is listed as ``Peak Intensity r.m.s.'' in
Table~\ref{tabclumpobs}.

The Hanning smoothed (over 3 channels) velocity profiles for 3 of these
enhancements are presented in \S~\ref{resultsenhance}. The spatial region of the
enhancement was selected via a rectangle and its profile plotted for a velocity
range larger than the range of the diffuse cloud component. We compared these
profiles to those produced by encompassing the diffuse emission of a cloud
component in a polygon which avoided the enhancements, confirming that the
velocity range of diffuse gas was larger than the range of each enhancement.
Also this comparison, along with visual examination of the cube, indicated which
velocity peak in the profile is associated with each enhancement. The FWHM and
the systemic velocity were measured for these narrower peaks using only the
region of the profile above the host's mean emission value.
The uncertainties in the enhancements' FWHM (a channel width, before smoothing,
of about 6 \kms) and central velocities reflect the uncertainty in visually
selecting velocity features.  Note that the C1 enhancement has 2 peaks that
blend around the 50\% level, so we do not quote a FWHM for this candidate. We
adopt as its systemic velocity the channel in which C's main features appear
simultaneously.  These characteristics are also presented in
\S~\ref{resultsenhance} and listed in Table~\ref{tabclumpobs}.

In order to measure the emission from an enhancement that does not include
emission from its host, we first created subcubes spanning only the FWHM
velocity range of the enhancement. We next subtracted from each channel of this
cube the
mean value of the emission due to the diffuse component of the cloud and
subsequently constructed integrated intensity maps, using the 2 different
approaches described in the previous section. The {\tt moment} maps are presented in
\S~\ref{resultsenhance}. 

Since the enhancements are only slightly larger than the dimensions of the
synthesized beam, the emission in each apparent enhancement was fit (using {\tt
  imfit} in {\sc Miriad}) with a Gaussian having the width of the
point-spread-function of the synthesized beam.  Using this fitting routine
allowed us to avoid nearby emission that would be measured within visually
selected polygons. With respect to uncertainties, we note that measuring B1 and C2 in integrated intensity maps formed from the
blanked cube increases the flux density by only 1-3\%. Instead our uncertainty
in the 
flux density of the enhancements, of less than 0.06 Jy$\times$\kms, is estimated
from the uncertainty in velocity (i.e. the selection of the number of channels
for the subcube) plus the 7\% increase that would occur if a polygon were used
for the fitting.  However we note that if the candidate enhancement's flux
density should include the emission currently attributed to the host cloud, then
its flux density could be larger by a factor up to 1.7.
 
This integrated flux density was used to calculate the amount
of HI mass in emission of each enhancement (e.g. equation $8.24$ in
\cite{1998gaas.book.....BinMer98}). Including the  uncertainty in H$_o$
generates
an uncertainty in mass of $\leq$ 29\%.

Although the position of an enhancement is the average of a few different
analysis techniques, to calculate its uncertainty we use the size of the
synthesized beam divided by the signal-to-noise. This results in a mean
uncertainty for the set of candidates of 26$^{''}$.

\section{Presentation of Measurements and Results}
\label{results} 

The measurements, analysis and results presented in this section are
used to consider  star formation possibilities within \cld (\S~\ref{derived}), 
for comparisons of \clds with other HI clouds
(\S~\ref{compare}), and in the discussion of the origin and fate of
\clds (\S~\ref{discussion}).

\subsection{Tracing \clds within the NGC~3256 Group}
\label{detectcloud}

\subsubsection{Radio data}
\label{detectradio}

Fig.~\ref{HIoverlay} shows \Cld's location in the NGC~3256 Group with the ATCA
HI emission overlaid on the Digitized Sky Survey\footnote{Based on photographic
  data obtained using The UK Schmidt Telescope. The UK Schmidt Telescope was
  operated by the Royal Observatory Edinburgh, with funding from the UK Science
  and Engineering Research Council, until 1988 June, and thereafter by the
  Anglo-Australian Observatory. Original plate material is copyright © the Royal
  Observatory Edinburgh and the Anglo-Australian Observatory. The plates were
  processed into the present compressed digital form with their permission. The
  Digitized Sky Survey was produced at the Space Telescope Science Institute
  under US Government grant NAG W-2166.}.  The velocity field of the NGC~3256
group and \clds is shown in Fig.~\ref{velfldoverlay}.  These images roughly span
700 \kms\ and 39\arcmin\ $\times$ 43\arcmin\ but do not show the whole group of
galaxies.

NGC~3256C, NGC~3256, and NGC~3263 (labelled in
Fig.~\ref{HIoverlay}) have clearly been disturbed by galaxy-galaxy interactions.
Predominant in these data is NGC~3263, which we determine has an HI mass of 2
$\times\ 10^{10}\ \mo$. It is noticeable in Fig.~\ref{HIoverlay} that the tidal
tail of NGC~3263 extends to the {\em east} of that galaxy, while \clds is
projected onto the plane of the sky to the {\em west} of NGC~3263. In
Fig.~\ref{velfldoverlay} NGC~3263's western velocity is quite distinct from that
of \cld. At NGC~3263's eastern side the velocity is 3309 \kms\ while its western
side's velocity is 2668 \kms.  Indeed \cld's velocity range of 2786 \kms\ to
2938 \kms\ (Table~\ref{combinedtable}) is similar to the eastern tidal tail of
NGC~3256.

Although they are usually not included in the Group in previous studies, we also
consider ESO 263-G044 and ESO 263-G046 to be members of the NGC~3256 Group.  The
average of NGC~3263's systemic velocity (2989 \kms) and that of NGC~3256 (2820
\kms; \citealt{2003AJ....125.1134EngNor}) is roughly 2900 \kms, which we adopt
for the moment to represent the central velocity for the group.  The HI profiles
of ESO 263-G044 and ESO 263-G046 give, with uncertainties less than a channel
width, systemic velocities of 3064 \kms\ and 3067 \kms, respectively. Thus,
assuming a moderate group velocity dispersion ($\leq$ 300 \kms), they 
are likely part of this system of galaxies.

This inclusion in the Group is supported by a possible bridge of HI that appears
between NGC~3263 and ESO263-G044; please see Fig.~\ref{bridge}.  With respect to
velocity continuity, the middle of the bridge appears in almost all channels
from 3051 \kms\ to 3164 \kms\ while there is also an apparent trend for material
to emanate from ESO263-G044 (around $\alpha$ = 10\hr 29\mn 27.7\st, $\delta$ =
-44\degr 15\arcmin 58\arcsec) at the lower velocities and attach to NGC~3263 at
the higher velocities (around $\alpha$ = 10\hr 29\mn 23\farcs1, $\delta$ =
-44\degr 09\arcmin 48\arcsec). An integrated intensity map gives a conservative
column density of $\rm 3 \times 10^{19} \ atoms \ cm^{-2}$ while a
channel-by-channel calculation gives twice the value.  The integrated flux
density is 0.78 Jy $\rm \times\ $ \kms.  In \S~\ref{comparediffuse}, where we
discuss the origin of \cld, we note that this bridge is indicative of an
interaction between NGC~3263 and ESO263-G044.

With respect to \clds itself, we have designated the most northerly structure
component A and it arcs from about  $\alpha$~=~10\hr 28\mn 41\st,
$\delta$~=~-44\degr 03\arcmin 40\arcsec\ to $\alpha$~=~10\hr 28\mn 18\st,
$\delta$~=~-44\degr 03\arcmin 50\arcsec (J2000), with a length of about 40
kpc. (We avoided measuring features to the east since they may be part of
NGC~3263.)

While A is quite diffuse, component B appears to consist of a resolved
amorphous structure enclosed in a larger diffuse elliptically shaped
cloud. We refer to the more resolved structure as ``interior'' and the
larger envelope as ``exterior''; these structures are labelled in the schematic
Fig.~\ref{schematic} which can be 
compared to the emission at velocities 2892.0 and 2911.9 \kms\ in
Fig.~\ref{velfldmosaic}. The exterior structure is centred around
$\alpha$~=~10\hr 28\mn 28\st, $\delta$~=~-44\degr 08\arcmin 40\arcsec\
and covers a region at least 67 kpc wide.  The interior structure
contains 2 intensity enhancements.

Component C runs diagonally for almost 139 kpc from northeast at
$\alpha$~=~10\hr 28\mn 50\st, 
$\delta$~=~-44\degr 08\arcmin 30\arcsec\
(beginning at B's east most side) to the southwest at $\alpha$~=~10\hr
28\mn 6\st, \\ $\delta$~=~-~44\degr 18\arcmin 19\arcsec.  It's projected
width is typically about 20 kpc and spans 47 kpc at its widest part.
Component C also has 2 intensity enhancements. 
We do not
report a FWHM for C since it has multiple velocity features. 

The observed and derived characteristics of the diffuse components A, B, and C are
given in Table~\ref{combinedtable}, while the intensity enhancements are 
described in \S~\ref{resultsenhance}.
From Table~\ref{combinedtable} it can be noted that the full velocity range
covered by \clds is 152 \kms and each component's velocity dispersion, \eg
FWHM/2.35,
is greater than 
the typical turbulent velocity (10 \kms) in the Milky Way's ISM. 
The largest difference between the systemic velocities of the components
is 61 \kms.  Although some portions of C are blueshifted relative to
the other components, there are no velocity signatures that clearly
indicate that \cld, or any component of \cld, is rotating, expanding
or collapsing. Of course if our line of sight is perpendicular to the
cloud's motion, we would not observe rotation or flows if these were occurring
along the plane of the sky.

\subsubsection{Optical data}
\label{detectoptical}

In the \ha emission data from the SuperCosmos I Sky Survey
\citep{2001MNRAS.326.1279Hambly} the region of \clds has the same statistical
characteristics as the other empty regions of the sky. Outlined on
Fig.~\ref{zoomcloud} is the annular region observed with the TAURUS Fabry-Perot
on the Anglo-Australian Telescope (\S~\ref{opticalobs}).  \ha emission was not
detected  in these data either. The 3$\sigma$ level non-detection corresponds
to 30 - 60 mR, or a surface brightness on order of $10^{-19}$ erg cm$^{-2}$
s$^{-1}$ (\S~\ref{opticalobs}), which indicates that the portion of the cloud
within the annulus plotted on Component B is unlikely to be ionized by the UV
radiation field of NGC~3263 or by strong shocks that would be created by
colliding gas. Although the cosmic background does produce very weak levels of
\ha emission \citep{1994ApJ...437L..95Bland94}, we would have needed about 16
hours and photometric conditions to reach these depths using the Fabry-Perot
staring technique.

Since this region was selected using preliminary HI data, more \ha
observations would be required in order to cover other regions of the
cloud; this is discussed in \S~\ref{uvmodel}.

A preliminary search for 
starlight in the region of \clds
was undertaken with CCD imaging at the
40-inch Siding Spring telescope.  The sky background in the V passband was 20.84
mag arcsec$^{-2}$, while the peak in the cloud was 20.82 mag arcsec$^{-2}$.  This
2\% increase in surface brightness corresponds to only
4 $\rm \sigma$ of the sky background so is not a significant detection. This
image indicates that there is no evidence for starlight in \clds at a surface
brightness exceeding about 3 $L_\odot$ pc$^{-2}$.
Although the position of \clds is at low
Galactic latitude, supplemental observations at infrared wavelengths could prove
helpful in confirming this.

\subsection{Delineating the Enhancements in \cld}
\label{resultsenhance}

We find 4 well-defined enhancements in the emission when analyzing the HI cube
(Table~\ref{atcalog}) using the criteria described in \S~\ref{obsenhancements}.
(That section also presents measurement methods and uncertainty estimates for
measurements presented in this section.)  Although each map in
Fig.~\ref{velfldmosaic} combines 3 channels from the HI data cube, the
enhancements in cloud component C are clear at 2792.7 \kms\ while the
enhancements in component B are evident at 2892 \kms.  The enhancements are also
evident at these velocities in the profiles (\S~\ref{obsenhancements}) presented
in Figure~\ref{enhanceprof}.  These 3 profiles include some of the diffuse cloud
emission which is spread over a broader velocity range
(e.g.~Table~\ref{combinedtable}). However the width of the narrow peak, measured
above the mean HI intensity of the host diffuse component, is consistent with
each enhancement's velocity range as determined by visually examining the cube.

We are 
motivated to distinguish between the host's emission and the enhancement's
emission in order to explore whether star formation occurs in the enhancement
candidates in \clds as it does in molecular density enhancements embedded in
diffuse clouds (\S~\ref{discussmass}).  
Thus we subtracted the mean
flux density of each diffuse host component and created intensity maps
integrated over the FWHM of the enhancements (\S~\ref{obsenhancements}). The
emission enhancements are clearly seen in these maps (Fig.~\ref{plotclumps}).
Their positions, labeled on Fig.~\ref{zoomcloud}, show that they do not coincide
with 2MASS infra-red sources.

The observed characteristics of the enhancements are listed in Table~\ref{tabclumpobs}.  For
example, the peak column density of each of the enhancements is on order of
10$^{20}$ atoms cm$^{-2}$.  The listed integrated flux densities 
have been converted into {$\rm {\cal M}_{HI}$}. These HI emission
masses are of order 10$^{8} \ \mo$
and are listed in Table~\ref{tabclumpsderive} along with the enhancements' 
velocity dispersions, which are
$\rm \sigma_{o}<$ 20 \kms.

C2 has the highest integrated flux density, and hence HI mass, 
and the largest velocity
width of the 3 measured enhancements. 
Interestingly, projected onto the plane of the sky
about 30\arcsec\ west of the peak of C2, and within this enhancement's
bounds, is an HII galaxy, WPV060, cataloged by Winkler et
al.~(1994)\nocite{1992A&AS...94..103Winkler}. Although barely resolved
in SuperCosmos data, this galaxy (also known as 2MASX
J10282575-4416172) has an estimated diameter of 7.4\arcsec\ at an
isophote of 20.0 K-mag arcsec$^{-2}$ (NED)\footnote{This research has
made use of the NASA/IPAC Extragalactic Database (NED) which is
operated by the Jet Propulsion Laboratory, California Institute of
Technology, under contract with the National Aeronautics and Space
Administration.}.  However the 
comparison in the paragraph below, of its optical spectrum
(described in \S~\ref{opticalobs}) with ATCA HI emission,  shows that
WPV060 does not appear, at least at the resolution scales 
of the HI cube listed in Table~\ref{atcalog}, to
be associated with the HI of enhancement C2.

The heliocentric velocity of WPV060, as measured from 5 emission lines
(H$\alpha$, [NII], [SII]: see Fig.~\ref{haspec}) in our optical spectrum, is
$2675 \pm 6$ km s$^{-1}$ (\S~\ref{opticalobs}), while the HI profile of C2 peaks
at $2815 \pm 3$ km$^{-1}$.  The emission lines are unresolved by the
optical spectrometer.  The integrated HI intensity map ( Fig.~\ref{haspec}) shows
no significant HI emission associated with WPV060 within $\pm 50$
km$^{-1}$ of its optical systemic velocity. While WPV060 is likely to
be a member of the NGC 3256 group, and one could speculate that the gas of
WPV060 could have been stripped, there is no conclusive evidence that it and C2
are related. 


\section{Derived Characteristics Relevant to Forming Star Clusters}
\label{derived}

We wish to explore \cld's potential to form stars within the candidate 
density enhancements. Therefore in this section we focus on deriving mass
and volume estimates, which in turn constrain values of density and pressure.
We consider a scenario in which the cloud components are in pressure
equilibrium and, if so, assess whether the enhancements may collapse to form
stars by representing them as Bonnor-Ebert spheres.

\subsection{HI Mass}
\label{derivedmass}
The total HI mass estimate for \clds -- [3 to 5] $\times \ 10^9 \mo$ -- is on
order of the HI content of normal Sc galaxies (\eg
\citealt{1980ApJ...242..903Bothun}).  The lower limit for this estimate comes
from our analysis of the ATCA HI cube (Table~\ref{atcalog}). That is, the sum of
the integrated flux densities of the 3 diffuse components
(Table~\ref{combinedtable}), which is 10 $\pm$ 3 Jy \kms, shows that the cloud
contains at least 3 $\times 10^9 \cal M_{\odot}$, with an uncertainty of 33\%
(\S~\ref{21measuremom}).
For the upper limit, we use the Parkes Telescope single-dish data to compensate
for the emission not detected by an interferometer.  \Clds was detected in 3
pointings \citep{eng94} and Fig.~\ref{parkesprofile} shows the velocity profile
at the pointing $\alpha$~=~10\hr 28\mn 27\fs7, \break $\delta$~=~-44\degr
09\arcmin 26\arcsec (J2000). Comparison of the Parkes profiles with the ATCA
data indicates that the narrow Gaussian component in the profile is associated
with \cld, rather than NGC~3263.  Using the area under the peak of this
component in Fig.~\ref{parkesprofile}, between 2760 and 2975 \kms, generates an
upper limit for the total emission of 15 $\pm$ 2 Jy \kms\ and a
corresponding HI mass for the cloud of about 5 $\times\ 10^9 \mo$.  The mass
estimate is also comparable to the combined mass associated with HI emission in both
tails of NGC~3256, $\rm \sim4.7 \times 10^9 \mo$
\citep{2003AJ....125.1134EngNor}.

The mass in HI emission for each enhancement alone, listed in
Table~\ref{tabclumpsderive}, is about 1 $\times \ 10^8$. (Recall that this
estimate is conservative since we subtracted emission which could be attributed
to the diffuse host cloud.)  \nocite{2003AJ....125.1134EngNor}English et
al.~(2003) argue that, if star formation has only 10\% efficiency, the amount of
gas required in order to form massive clusters of stars, such as globular
clusters, is on the order of $\rm 10^{7 \pm 1 }\mo$. Thus the enhancements' HI
mass is at the high end of this mass range.

\subsection{Pressure Equilibrium} 
\subsubsection{Deriving Pressure} 

\label{analysisderive}

The pressure of the diffuse HI gas component, required for our Bonnor-Ebert
sphere analysis in the following section, can be determined from \beq \rm P =
\rho \sigma^2\eeq where $\rho$ is the volume density and $\sigma$ is the
velocity dispersion.  Although the mean density of a region of atomic hydrogen
is simply $\rho = \cal M_{\rm HI}\ / \ {\rm volume}$, we need to adopt a spatial
scale along the line of sight (LOS).  

We measure the pressure in the diffuse gas of A and C by selecting in each
component a representative rectangular area, avoiding enhancement candidates. In
component C, we additionally avoid the diagonal structure at redshifts $\geq$
2825 \kms.  For the LOS axis we took the average of the region's length (e.g.
major axis) and width (e.g. minor axis).  The region in C is 20 kpc $\times$ 16
kpc, encloses 1.06 $\times 10^8 \mo$ of HI, has an uncertainty in the LOS axis
of 9\%, the standard deviation in $\sigma$ is 11\%, and the combined uncertainty
in pressure is at least 40\%. In component A the region is 26 kpc $\times$ 49 kpc,
contains 1.56 $\times 10^7 \mo$ of HI, and the uncertainty in the LOS axis is at least
24\%. The uncertainty in A's pressure may be larger than 50\% which is indicated
by the difference between the velocity range determined by visual inspection of
the HI cube and the FWHM measured from the intensity profiles
(\S~\ref{21measuremom}).

Cloud component B is more complex. \S~\ref{detectradio} describes the
interior and exterior structures of B (also see Fig.~\ref{schematic})
and the values of the parameters used to derive mass and pressure are
presented in Table~\ref{combinedtable}. For each structure in B we
calculate the pressure for two simple models. One is that of an
oblate-shaped cloud (i.e. half the length of the LOS axis = the
semi-major axis) and the other for an prolate-shaped cloud (i.e. half
the length of the LOS axis = semi-minor axis). We then average these
estimates together.  To determine the mass and volume of the exterior
component alone,
the mass and volume of the interior component of
B were subtracted from those values for the whole cloud.  Combining
the uncertainties from the flux density (i.e.~the dominant uncertainty),
the measurement of the axes (large for the LOS axis), the FWHM, and
the distance, there is an uncertainty in each structure's pressure of
at least 43\%.

The pressure  (P/k) of each diffuse component is presented in
Table~\ref{combinedtable}; They are normalized by the Boltzmann
constant
for comparison with
pressures in theoretical studies.  These pressures, which range from
5 to 100 K $\rm cm^{-3}$,
are used in
\S~\ref{discussmass} where we calculate the Bonnor-Ebert mass of each
enhancement candidate.

\subsubsection{Pressure Equilibrium Discussion} 
\label{discusspressure}

If the diffuse part of the cloud were in pressure equilibrium, then we could
employ virial theorem arguments to estimate the masses (independent of the mass
in HI emission) required if the enhancements are about to form stars
(\S~\ref{discussmass}).  As pointed out in \S~\ref{detectcloud}, component B
appears to consist of a resolved structure interior to a more diffuse, exterior
shell; see Fig.~\ref{schematic}.  This is analogous to the well-explored
theoretical scenario of a molecular cloud surrounded by the ISM's HI. Applying
the analogy to our data, if the pressure in the interior HI structure is at
least equal to that in the exterior HI shell then these 2 structures could be in
pressure equilibrium.  Although the pressures of these structures, listed in
Table~\ref{combinedtable}, differ by 40\%, the error in the estimate of the
exterior pressure is at least 43\% (\S~\ref{analysisderive}). Therefore we can
consider the possibility that the interior and exterior diffuse structures are
in pressure equilibrium. Subsequently we use the average of the interior
structure's pressure and the exterior shell's pressure as the value for the
ISM's pressure in cloud B.



Although we then assume that \clds is in equilibrium for subsequent
calculations, the clumps and irregular morphology indicate that this
may not be the case.  A dynamical crossing time in component B,
estimated using the radius divided by the velocity dispersion, is 1-2
Gyr. If this is a primordial cloud then there has
been sufficient time for virialization. However the age estimate of
the tidal tails of NGC~3256 \citep{2003AJ....125.1134EngNor} is 500
Myrs.  Hence the cloud may not yet be in equilibrium
if the cloud is
the remnant of the galaxy-galaxy interactions which also precipitated
the formation of NGC~3256. 

\subsection{Bonnor-Ebert Mass and Star Formation in the Enhancements}
\label{discussmass}

Again applying the molecular cloud analogy to our data, the enhancements, which
appear to be distinct velocity features in velocity profiles
(Figure~\ref{enhanceprof}), would be analogous to molecular cores. In this
section we consider whether, given the conditions in the surrounding diffuse
gas, there is sufficient mass in the enhancements to cause star formation.  This
analysis uses the velocity dispersion of each enhancement ($\sigma_o$) and the
pressure of its host diffuse component. That is, we assume that the enhancements
are in pressure equilibrium with the HI in which they are embedded and hence
their surface pressures ($\rm P_s$) are equivalent to the pressures in the
diffuse cloud components.  The pressure is $\rm P/k \sim 72\ K\ cm^{-3}$
for the diffuse component B
(\S~\ref{discusspressure})
and 
$\rm P/k \sim 39\ K\ cm^{-3}$ for component C
(Table~\ref{combinedtable}).  
Including uncertainties in their measurements, the 
pressure in B is $\rm P/k \sim 32 - 100\ K\ cm^{-3}$
and C has $\rm P/k \sim 23 - 55\ K\ cm^{-3}$.
Although each enhancement contains a
sufficient HI gas mass to populate a globular cluster if it were
converted to stars, these pressures are far short of the 
$\rm P/k = 10^6- 10^8\ K\  cm^{-3}$
proposed by \cite{1997ApJ...480..235elmef97} as a
requirement for the formation of such massive, bound structures.

However we can calculate, using the given pressure and velocity
conditions, the amount of mass that would be required by each
enhancement in order for it to form stars, regardless of whether they
are in a cluster.  To do so we assume that each embedded enhancement 
is an isothermal sphere on the verge of collapse (i.e.~a Bonnor-Ebert
sphere). We note that the enhancements, which are slightly larger than the
synthesized beam, may be approximated by spheres in these data
(e.g. Figure~\ref{plotclumps}).   \cite{1956MNRAS.116..351Bonnor} defines the radius of the
interface of an isothermal sphere with the ISM as 
\beq r_c = 0.49 (\frac{k\ T}{m\ G\ \rho_c})^{1/2} \eeq 
where m is the molecular weight and $\rm \rho_c$, the critical density, 
is the density of the sphere also at
the interface.  The relationship between the mass within this radius
and $\rm r_c$ is 
\beq M(r_c) = 2.4 \ \frac{k \ T}{m\ G} \ r_c \eeq
Substituting the first equation into the second, and using $\rm \rho_c = P_{s} /
\sigma_o^2$, and $\rm \sigma_o^2 = k \ T / m$, generates 
\beq M(r_c) = 1.18 \ \frac{\sigma_o^4}{G^{3/2} \ P_{s}^{1/2}} \eeq 
where $\rm \sigma_o$ is the velocity dispersion of the enhancement
(i.e.  FWHM of the enhancement in Table~\ref{tabclumpobs} divided by
2.35) and $\rm P_{s}$ = P of the ISM in the host component.  
If we had used  $\sigma_o^2 \ \times \ r_c  \ \times G^{-1}$ for the virial
mass,  we would be
forced to adopt the beam size as an upper limit for the unknown
diameter.  Thus it is noteworthy that the formulation above eliminates the
need to measure the intrinsic radius of an enhancement.

These derived masses are listed in
Table~\ref{tabclumpsderive}. Combining the largest uncertainties in
$\rm P_{s}$ and $\rm \sigma_o$ generates an uncertainty in the
Bonnor-Ebert mass of about 40\%.
This table also lists the ratio of the estimated Bonnor-Ebert mass to the HI
mass in emission.  Accounting for the mass range given by the uncertainties,
this ratio would need to be $\frac{\cal M\rm(r_c)}{\mh}\leq$2.2 for the HI mass
to be equivalent to the Bonnor-Ebert mass, and hence star formation to be a
possibility. However the $M(r_c)$ are 9-44 times larger than $\mh$. We note that
the HI mass measurement is a lower limit since we subtracted off a threshold
that we assumed to be associated with the host cloud emission. Therefore if some
of the emission attributed to the host belongs to the enhancements, then the HI
enhancement candidates could have almost twice as much emission
(\S~\ref{obsenhancements}) than listed. Taking into account that this assumption
also effects
the FWHM measurement, and therefore the value of $\rm \sigma_o$, would generate
the mass ratios of about 6-22.

The analysis above indicates that if these enhancements are isothermal spheres
composed solely of atomic hydrogen gas there is insufficient gas for star
formation to occur.  Using the values listed in Table~\ref{tabclumpsderive}, the
amount of gas is deficient by about an order of magnitude; alternatively, the
velocities within the enhancements are about 2 times too fast given the pressure
of the surrounding medium.  Another perspective is that the external pressure
would need to increase in order to cause the enhancements to collapse and form
stars.  The increase in external pressure required is at least a factor of 9 and
possibly as large as a factor of 400. The former estimate is based on B2
and attributes all of the observed emission to the enhancement, while the
latter factor uses the tabulated HI mass estimate for C2.


\subsection{The Ultra-Violet Radiation Field of NGC~3263}
\label{uvmodel}

The non-detection of ionized hydrogen noted in \S~\ref{detectoptical} indicates
that the portion within the TAURUS annulus is unlikely to be ionized by
NGC~3263. In the current section we explore whether other regions of \clds
could nevertheless be ionized. 

We have computed the ionizing radiation field from NGC 3263 by scaling
up the Galaxy model of Bland-Hawthorn \& Maloney
(\nocite{1999ApJ...510L..33BlandMal99}1999;
\nocite{2002ASPC..254..267BlandMal02}2002) by the ratio of the total
blue luminosities, such that $\rm L_{B(NGC 3263)}/L_{B(Galaxy)}$ = 10.
Without a detailed knowledge of the disk star formation history in
this highly inclined galaxy, $\rm L_B$ is a useful surrogate.  The
measured ratio of $\sim$5 has been doubled in order to generate a
face-on model which is appropriate for a disk-dominated galaxy
(see Driver et al 2007, Fig. 6). In our model, we adopt a total disk diameter of 20
kpc and a mean disk opacity of $\tau_{LL} = 2.8$ where $\tau_{LL}$ is
the opacity at the Lyman limit.

In Fig.~\ref{uvfld}, we show the distribution
of ionizing radiation above and below the galaxy plane.
The uniform dust distribution in the disk elongates the radiation
field along the minor axis. The contours show lines of
constant ionizing flux. These can be converted to the
expected emission measure in H$\alpha$ with the
following formula \citep{1999ApJ...510L..33BlandMal99}

\begin{equation}
\log(EM) = \log\phi - 3.38
\end{equation}
where the emission measure $EM$ is given in milliRayleighs (mR).  The contours in
Fig.~\ref{uvfld} are given in units of $\log\phi$. Thus, at a distance of 80 kpc
off the plane, the predicted ionizing flux is $\log\phi = 5$. This is sufficient
to excite an HI cloud to an emission measure of about 400 mR which can be
detected in a deep exposure with a tunable imaging filter
(e.g. \nocite{2003AJ....126.2185Veil} Veilleux et~al.~2003). The Fabry-Perot
staring method is able to improve on this by an order of magnitude
(\nocite{1994ApJ...437L..95Bland94}Bland-Hawthorn et al. 1994) which means that
most of the HI even far from the spin axis of the galaxy (where $\log\phi >
4.5$) would be ionized and therefore rendered visible to this technique. Assuming \clds is
at the same distance as NGC 3263, if this is found not to be true, then the
covering fraction of the gas, as seen from the galaxy, may be low. Alternatively our
argument could be affected by uncertainty in the NGC~3263 model due to scaling
up the Milky Way model, or the polar dependence of the radiation field is
stronger than indicated by our model
\citep{1999ApJ...510L..33BlandMal99}. 

However if our model is appropriate, the northern section of component C lies
long the $\log\phi = 5$ contour.  It has a column density of $\rm 4.58\ \times\
10^{19}\ atoms\ cm^{-2}$ which is almost half that of C's southern section.
Although observationally challenging, emission measures of order 40 mR
should be detectable from this gas if it falls at the distance of NGC 3263. We
note in passing that the expected H$\alpha$ emission measure from the cosmic ionizing
background is an order of magnitude smaller which falls close to the systematic limit
of the technique (Bland-Hawthorn et al 1994).

\section{Comparison of \Clds with Other Clouds}
\label{compare}

\subsection{\Cld's Diffuse Component, Isolated Clouds, and Tidal Features} 
\label{comparediffuse}

Summarizing the characteristics in Table~\ref{combinedtable}, \cld's column
density range is $\rm 10^{19-20}\ atoms\ cm^{-2}$, HI mass range is $\rm
10^{8-9}\ \mo$ and FWHM of 41-97 \kms; its extent is 100 $\times$ 175 kpc.  With
respect to apparently isolated clouds, as pointed out by Ryder et
al.~(2001)\nocite{2001ApJ...555..232Ryder}, \clds has a similar mass and extent
as HIPASS J0731-69. However \cld's extent is larger than FCC 35's HI companion
cloud (13 kpc; \citet{1998AJ....115.2345Putman}) and the HIJASS J1021+6842 cloud
in the M82 Group (30 kpc; \citet{2005ApJ...627L.105WalSkil}). Since these
isolated clouds are dissimilar from each other, we cannot draw strong
conclusions about \clds based on these comparisons.  Therefore, in this section,
we focus on comparing the column density of the diffuse components of \clds with
tidally produced features.

\Clds is morphologically less contiguous and less regular in velocity than even
very unusual kinds of tidal debris (\eg IC~2554 \citep{2003MNRAS.339.1203Kor})
produced by galaxy-galaxy interactions and mergers. In terms of other
characteristics, \clds is most similar to the debris in clusters of galaxies,
such as the plume near NGC~4388 and the HI tail from NGC~4254 in which the
putative dark galaxy VirgoHI 21 (\eg \citet{2007ApJ...670.1056Min}) is
embedded. That is, the plume near NGC~4388 has $\rm 10^{20}\ atoms\ cm^{-2}$, a
few $\rm \times 10^{8}\ \mo$, a FWHM of 100 \kms, and an extent of 110 $\times$ 25 kpc
\citep{2005AA...437L..19OosVG}.  In the ALFALFA data, NGC~4254's tail has $\rm 6
\times 10^{8}\ \mo$ and an extent of roughly 250 kpc
\citep{2007ApJ...665L..19Haynes}.

With respect to tidal debris in groups of galaxies, we can compare \clds with
tidal features in the field of view of our HI data on the NGC~3256 Group.  For
example, a possible bridge of HI material, 67 kpc long, appears 
to connect NGC~3263 and ESO263-G044 (\S~\ref{detectradio}, Fig.~\ref{bridge}).
The bridge's column density, $\rm 3-6 \times 10^{19} \ atoms \ cm^{-2}$, is
similar to \cld's diffuse components. However the integrated flux density
corresponds to HI mass of roughly 7 $\rm \times\ 10^6 \mo$, which is lower than
component A by an order of magnitude.

While the combined mass in both tails of NGC~3256 is comparable to the total
mass in \cld, the eastern tidal tail's column density, $\rm 1.6 \times 10^{20}\
atoms\ cm^{-2}$ is larger than that of any individual component in \cld. (The
value for the tail is determined from the HI cube presented in the current paper
and has an uncertainty of 5\%.)  Note that the fragment north of NGC~3256 (called
Fragment A2 in \citealt{2003AJ....125.1134EngNor}) has a column density, in the
current data, of $\rm 1.7 \times 10^{19}\ atoms\ cm^{-2}$, similar to
component~A.

\subsection{\Cld's Enhancements Versus Apparently Isolated Clouds}
\label{compareclumps}

\Cld's few enhancement candidates have a peak column density on the order of
10$^{20}$ atoms cm$^{-2}$, masses of about 10$^{8}\mo$, FWHM of 20 to 44 \kms,
and diameters of about 14 kpc (see Table~\ref{tabclumpobs},
Table~\ref{combinedtable}, and Table~\ref{tabclumpsderive}). Although there are
observational limits due to different angular resolutions (noted at the end of
this section), we contrast these enhancements with observations of various kinds
of HI clouds to see if similarities or differences can help illuminate \cld's
origin or fate (\S~\ref{fate}). 

In general we find that \cld's enhancements cannot easily be categorized with
isolated HI clouds but instead differ from them in a variety of ways.  For
example the clumps in HIPASS J0731-69 each have almost as much HI mass as \cld's
enhancements ($\rm \sim 5\ \times\ 10^{7}\ \mo$; \citet{2004ryderkor}) yet the
peak column density is less 
(1.3 $\rm \times \ 10^{19}$ atoms cm$^{-2}$;
\citet{2001ApJ...555..232Ryder}). The peak column density is also less in the HI
companion to FCC 35 (2.0 $\rm \times \ 10^{19}$ atoms cm$^{-2}$;
\citet{1998AJ....115.2345Putman}).  The HI companions to HII galaxies detected
by \nocite{1993AJ....105..128Taylor,1995ApJS...99..427Taybrin} Taylor~\etal
(1993, 1995) have some characteristics that are similar to those of \clds
enhancements; peak column densities ranging from 10$^{19}$ to 10$^{21}$ atoms
cm$^{-2}$, HI masses from $\rm 10^8 \ to\ 10^{10}\ \mo$ and diameters as small
as a few kpc to a few tens of kpc.  However, the companions'
Half-Width-Half-Maximum range from 30 to 240 \kms\ while \cld's enhancements'
$\rm \sigma_o$'s range from 13 to 19 \kms.  However there are optically unseen
clouds in the Virgo cluster which have roughly $\rm 10^8 \ \mo$ and one cloud in
particular consists of 5 clumps strewn across 200 kpc; a few of these clumps
have FWHMs of 50-60 \kms\ \citep{2008IAUS..244...93Kent}.

Convincing detections of stellar populations in HVCs associated with our Milky
Way Galaxy remain elusive (e.g.  \citealt{1983ApJ...273L...1Schneider},
\citealt{2002AJ....124.2600W}, \citealt{2002ApJ...574..726Simon},
\citealt{2006ApJ...640..270Simon}).  Therefore HVCs are the most nearby HI
clouds available for comparison with the enhancements in \cld. Some rare compact
HVCs (CHVCs) might appear to be potential analogues to \cld's {\em enhancements}
since CHVC have FWHMs of 25-35 \kms\ \citep{2002A&A...392..417DeHeij} and can
have peak column densities of $\rm 10^{20} \ cm^{-2}$
\citep{2002AJ....123..873Pmarymany}.  However distances to some HVC indicate
that the population resides within tens of kpc from the Galactic plane
(\eg\citealt{2006A&A...455..481Kalberla} and references mentioned within;
\citealt{2008ApJ...672..298Wakker}). Hence most CHCVs, which typically have an
angular size of $\sim$~50 arcmin, would be much smaller than the
enhancements. To set a distance limit, a CHVC would need to reside at 1 Mpc from
the Milky Way for the corresponding linear diameter to be 15 kpc and the CHVC to
be comparable to \cld's enhancements.


Some non-compact HVCs have density enhancements surrounded by HI with column
densities around $\rm \sim 2 \ \times \ 10^{19} \ cm^{-2}$
(\citealt{2002A&A...391..159DeHeij}, \citealt{2006A&A...455..481Kalberla}), which
is comparable to
the column density of the surrounding diffuse component of \cld. So we can
consider whether the diffuse major components of \clds are similar to these
HVC. Peak column densities in HVC complexes can be on order of $\rm \sim 10^{20}
\ cm^{-2}$, e.g. complex A \citep{1991A&A...250..484WakSch} and the HIPASS HVCs
\citep{2002AJ....123..873Pmarymany}.  However these particular HVCs have typical
linewidths of only 35 \kms, whereas individual pointings on cloud component B
are all greater than 41 \kms\ and often greater than 80 \kms.
As a specific example, Complex A's envelope has a linewidth of about 25 \kms,
the FWHM of its cores is only 10 \kms\ \citep{1997ARA&A..35..217WakWoer}, and
its mass is only 10$^{5-6} \mo$ \citep{1999Natur.400..138VanWoer}. Work by
\nocite{2006A&A...455..481Kalberla}Kalberla \& Haud (2006) find the cores of the
HVC typically have velocity dispersions of a few \kms, although the $\sigma_{o}
>$ 10 \kms\ of \clds enhancements does not appear excluded.  However they show
that the envelops in HVC complexes have velocity dispersions of about 12 \kms,
compared to our estimate (FWHM/2.35) of 41 \kms, associated with the whole
diffuse component of Cloud B (Table~\ref{combinedtable}}).  Thus the {\em diffuse
components}, in particular, of \clds are evidently dissimilar to the envelopes
of the HVC complexes.

We note that there are caveats to our comparisons. The use of different beam
sizes and velocity resolutions can play a role in generating discrepancies when
comparing extragalactic and Galactic clouds.  The observational parameters are
also convolved with different physical scales since the clouds are at different
distances. Measurements of FWHM are problematic due to, for example, multiple
peaks in the profiles and beam smearing of the velocity gradient.  Additionally
peak column densities are defined and measured differently in different papers.
Nevertheless by doing these comparisons we find that \cld, rather than clearly
falling into a specific category of extragalactic HI cloud, instead contributes
to the variety of types of extragalactic HI clouds.
Also \cld's enhancements may be similar only to CHVC and then only in the
unlikely case that CHVC are at large distances from the Milky Way.

\section{Discussion}
\label{discussion}
\subsection{Origin of \cld}

\Cld, as delineated by our ATCA data, contains a few $\times 10^9 \mo$
of HI which is comparable to that in an Sc galaxy. However if this
cloud contains stars or ionized hydrogen gas, these components are not
obvious in our optical data. Therefore, since galaxy formation is not
expected to be 100\% efficient, the possibility exists that \clds is a
primordial cloud, an example of a building block in hierarchical
galaxy formation scenario. Although various extragalactic clouds have
been detected and characterized, the rarity of bona fide primordial
clouds (\S~\ref{introclouds}) means that their morphology, column
density, and velocity behaviour is unknown. Thus primordial cloud
characteristics can not be used to support this origin for \cld.

Since \clds resides in a group environment in which most galaxies show
tidal disturbances (e.g. Fig.\ref{HIoverlay}), it is more likely that
\cld's origin is 
tidal as well. A few properties of \clds also suggest a
tidal origin. For example, although \clds is more tenuous, its total mass is
similar to the amount in the tidal tails of NGC~3256.  Also its column
density is similar to NGC~3256's fragment A2
(\S~\ref{comparediffuse}). Therefore the intention of this section is
to review a few relevant interaction scenarios, in light of the
properties of \clds and \cld's environment, in order to assess which
of these would be most relevant to \cld's origin.

\cite{2005MNRAS.357L..21BekKor} have numerically studied the effect on
individual gas-rich galaxies due to the gravitational potential of a group of
galaxies. Their model of a low surface brightness late-type galaxy orbiting the
group's centre generates a starless Leo ring-type of HI structure within about 6
Gyr. The outer disk gas, which in this galaxy extends 5 times that of the
stellar disk, is stripped off of the parent galaxy and enroute, at $\sim$4 Gyr,
the tidal structure is
similar to that of HIPASS J0731-69. The observed morphology is the result of the
range of HI column densities. That is, the structure is contiguous but with an
inhomogeneous density such that only high density regions of the tidal arcs are
revealed using current radio arrays.  This scenario is appealing since
qualitatively the structure at 4 Gyr is suggestive of the morphology of NGC~3256
combined with \clds (compare our Fig.\ref{HIoverlay} and Fig.\ref{velfldoverlay}
to Fig.~1 in \cite{2005MNRAS.357L..21BekKor}). However this picture is probably
misleading for a number of reasons. In the model the parent galaxy does not form
{\em stellar} tidal tails, which NGC~3256 does exhibit, even in the Digitized
Sky Survey (e.g. Fig.\ref{HIoverlay}}).  Also the observed morphological
features of NGC~3256, including these stellar tails, can be more simply accounted
for by a major merger of 2 galaxies (e.g.~\citealt{2003AJ....125.1134EngNor}).
Additionally the model shows that the parent galaxy resides outside the radius
of the resultant ring, implying that galaxies other than NGC~3256 could have
provided \cld's gas instead.

A drawback of this group potential simulation is that it is particular about the
characteristics of the parent galaxy as well as the distance from the centre of
the group potential.  That is, star formation will occur in the arcs drawn from
a more massive galaxy and rings may not form if a high surface brightness or
compact galaxy interacts with the group potential. We can relax these parent
galaxy effects by using an analogy with clusters of galaxies since we expect the
tidal forces in groups of galaxies to be analogous to those in clusters,
although more minor in affect.  We are also motivated by our assessment that the
type of object that \clds most resembles is debris within clusters
(\S~\ref{comparediffuse}).

As galaxies orbit in a cluster environment, they are not only subject
to the cluster potential itself, but also to interactions with
individual galaxies. Take, for example, pairs of galaxies that
interact in the outer regions of the cluster as they fall in
(e.g.~\citealt{1998ApJ...502..141Dubinski}).  Although tidal material
pulled from the outer disk of each galaxy remains bound to an
interacting system if the interaction occurs in the field, within a
cluster environment these tails are also subject to the tidal field of
the cluster. This causes the tails to be more extended and diffuse.
It also efficiently strips the tail material out of the galactic
potential, dispersing it throughout the cluster
(\citealt{2004IAUS..217..390Mihos}, \citealt{2004cgpc.symp..278Mihos} and
papers mentioned therein; \citealt{2005bekkorkil}).

In the case of the NGC~3256 group, most of the galaxies have peculiar
morphologies that are similar to those in isolated interactions
(e.g. Fig.\ref{HIoverlay}).  None are so pathological that they require a mass
like \clds to exist in order to explain their features.  As previously
mentioned, NGC~3256 has the morphology of an on-going merger of 2-3 galaxies
(e.g. \citealt{2003AJ....125.1134EngNor}, \citealt{2000AJ....120..645Lipari}).
Another example is NGC~3263 which has an extended arm (\S~\ref{detectradio})
that can be generated when a smaller galaxy has an inclined orbit to a larger
one (\eg\citealt{1986ApJ...309..472Quinn}).  A candidate satellite galaxy can
even be identified in our data; ESO263-G044 appears related to NGC~3263 by an HI
bridge (\S~\ref{detectradio}, \S~\ref{comparediffuse}).  Indeed NGC~3263's arm
extends in the opposite direction to \clds (e.g. Fig.\ref{velfldoverlay},
\S~\ref{detectradio}), confirming that \clds is superfluous for producing the
tidal features of NGC~3263.

The apparent simplicity of these interacting systems suggests we should examine
a scenario for the formation of \clds which is analogous to the process
described for interactions in clusters. Originally a typical interaction creates
tidal features as the interacting pair falls into the group's potential. The
tidal field of the group causes these features to be more extended and diffuse
than they would be in the field.  The group potential subsequently strips the
tidal features off of the parent. Thus \clds, which has an HI mass on order of
the content of normal Sc galaxies (\S~\ref{derivedmass}), could be stripped tidal
features. Additionally, since the tidal features, and hence much of the
interacting galaxies' HI gas, are removed in this process, it is difficult to
recognize, by its morphology alone, any particular peculiar galaxy in the group
as being the specific victim of an appropriate encounter.  There are a number of
candidates, within the large region of the association of galaxies, which are
undetected in HI (e.g. ESO263-G045, 263-G8, 317-G4, N3250B, N3318A, 264-G14
\citep{1991ApJ...383..467Mould}). However if we assume that the process that
created \clds occurred roughly when the tails of NGC~3256 were generated, then
the most likely candidate is the gas-poor peculiar galaxy NGC~3256c.  The
projected distance between NGC~3256c and \cld, and their velocity difference,
implies that they were in close proximity roughly 600 Myr ago while the
timescale for NGC~3256's pericentre encounter was 500 Myr ago
\citep{2003AJ....125.1134EngNor}.

An alternative origin would be ram pressure stripping by the putative hot gas
between the galaxies in the NGC~3256 Group.  The situation would be similar to
the case of the structure containing VirgoHI 21, which does have properties
similar to \clds (\S~\ref{comparediffuse}).  VirgoHI~21 was considered to be an
example of ram pressure stripping by Oosterloo \& Van Gorkom
(2005)\nocite{2005AA...437L..19OosVG}.  In this case one would be tempted to
interpret the vertical extensions from NGC~3263, such as the ``bridge'', and
\clds as stripped gas.  However the Oosterloo \& Van Gorkom argument was based
on the observation that VirgoHI~21's parent galaxy NGC~4254 had a truncated HI
disk. Since the disk of NGC~3263 is gas rich and extended, this instead suggests
that a galaxy such as NGC~3256c would have been the victim of ram pressure
stripping. Unfortunately for the ram pressure scenario as applyed to VirgoHI~21
recent, more sensitive
observations show that VirgoHI 21 is embedded in an extensive HI plume from
NGC~4254 \citep{2007ApJ...665L..19Haynes} and this plume has been successfully
modelled as a tidal tail that was expelled from NGC~4254 during the flyby of a
massive companion galaxy \citep{2008ApJ...673..787Duc}.  Noting also that ram
pressure stripping does not play a major role in the removal of HI from galaxies
in, for example, Hickson Compact Groups \citep{2008MNRAS.388.1245Ras}, we do not
think a ram pressure origin is likely for \clds which resides in a group rather
than a cluster.

To date we have not observed a stellar component or ionized gas associated with
\clds (\S~\ref{detectoptical}). This is not surprising given that the pressure
in \clds is too low to create new stars (\S~\ref{discussmass}). Additionally
older stars could be too diffusely distributed to be detected or the tidal
material may only come from the more loosely bound, metal-poor outer disks of
the parent galaxies. A caveat to the latter scenario is that \cld's large HI
mass, which is comparable to the total fraction associated with a normal Sc
spiral galaxy, may require more gas than the amount available in outer disk
regions. CO observations would prove useful for testing this, even though an
absence of molecular gas in \clds could be for a number of reasons besides the
explanation that \cld's material came from outer disk regions.  If, however, CO
were detected, it would suggest that the tidal field of the group helped dredge
up gas from the parent galaxies' inner regions, where CO tends to reside
(e.g.~\citealt{1991ARAA..29..581Y}).  This would be in contrast to an isolated
interaction which usually causes this inner gas to flow towards the galactic
potential, converting into stars (e.g.~\citealt{1994ApJ...431L...9M}) or forming
an AGN (e.g.~\citealt{2006APJS..166....1HopHern}).


Clearly a model of this group  would be needed
to explore the validity of this scenario in which tidal features are
stripped away by a group's gravitational potential. In particular, the
morphology, kinematic behaviour, and star formation possibilities need
to be explored.

\subsection{Fate of \cld}
\label{fate}

Regardless of the origin of \cld, currently it is experiencing tidal forces due
the NGC 3256 Group's tidal field. If a group member is in the proximity of \clds
its gravitation field may also be effecting the cloud.  Such forces would cause
the majority of the cloud to become even more diffuse.  However could the tidal
field(s) also cause the more dense, embedded HI enhancements to evolve into
gravitationally bound structures such as tidal dwarfs and globular clusters?
These density enhancements have masses that are comparable to stellar groups on
such scales (\S~\ref{derivedmass}).
However, \S~\ref{discussmass} indicates that these enhancements are not
Bonnor-Ebert spheres.  That is, there is insufficient external pressure to
induce the density enhancements into forming stars.  Indeed, for the observed
velocities, the enhancements could require another order of magnitude of mass in
order for an enhancement to collapse. (Although it is speculative, assuming a
typical mass ratio of H$_2$-to-HI for an Sc galaxy \citep{1991ARAA..29..581Y},
the contribution from H$_2$ is negligible.) Therefore it is not a surprise
that the HII galaxy WPV060 does not appear to be a tidal dwarf galaxy embedded
in \cld.  Also the FWHM velocities in the enhancements are 2-4 times the typical
turbulent velocity of ISM ($\sim$10 \kms) and it is hard to imagine how \clds
velocities could be reduced.  So unless an event occurs to increase the pressure
in each cloud component that hosts an enhancement it is unlikely that these
enhancements will be identified in the future with either globular clusters or
dwarf galaxies, tidal or otherwise.

A model of the fate of this cloud in the gravitational potential of the group of
galaxies would be as interesting as a model of its origin. If there were a
possibility of \clds collapsing, this would take at least 2 Gyr, ie.~the
crossing time. This process may be in competition with the time it takes tidal
forces to form a structure having this morphology; this timescale is 2 -- 4 Gyr
using the example of the \cite{2005MNRAS.357L..21BekKor} model.  If, instead of
collapsing, the tidal forces win out then perhaps the enhancements will become
stripped of their surrounding medium. If additionally the enhancements become
less dense they may be identified with the more typical HVCs that surround our
Milky Way, rather than the rare denser CHVCs described in
\S~\ref{compareclumps}.  This is of particular interest since it suggests an
additional origin for HVCs.

Interestingly, Bland-Hawthorn et al (2007) have recently explained the anomalous
H$\alpha$ emission along the full length of the Magellanic Stream in terms of a
shock cascade along it. As the Stream falls through the extended hot halo of the
Galaxy, clouds at the head of the Stream disrupt and shower cool material onto
the fast-approaching clouds moving in the same direction. These authors argue
that the Magellanic Stream is dissolving and the warm gas is raining onto the
outer halo of the Galaxy. This may be the eventual fate of \clds if it
is in the proximity of NGC~3263. The Fabry-Perot staring method of section
\ref{detectoptical} could illuminate whether this scenario  is valid for  \cld.

Alternatively, if HVC-like enhancements form and are not accumulated or
dissolved by a specific galaxy, they should fall towards the group's centre of
mass.  Then these ``cloudlets'' consisting of tidal material, rather than
primordial material, may mimic the distribution and motion of primordial
building blocks falling into the potential well of the global dark matter
distribution of the group of galaxies.  The fact that so few enhancement
candidates exist in \clds has the potential to help discriminate between these 2
scenarios.  That is, if too few in-falling objects are observed, compared to the
numbers predicted by cosmological simulations, the observed cloudlets are
unlikely to be part of individual primordial dark matter subhalos. Rather
they are more likely to be tidal debris from an evolved Vela Cloud-like object.


\section{Summary}
\label{summary}

A rare galaxy-sized intergalactic cloud, \cld, which resides in the NGC~3256
Group of galaxies contains, according to our ATCA and Parkes single-dish
telescope observations, an HI mass of 3-5 $\times\ 10^9 \mo$.  While it is less
than the mass in some of the galaxies in this group, this is comparable to the
gas content of a normal spiral galaxy.

Since interferometer observations filter out low column densities and
specific spatial scales, in ATCA data
 the cloud appears as 3 separate components containing 4
robust HI density enhancement candidates. We have not discovered a stellar
component nor \ha radiation in our preliminary optical
investigations. However if parts of \clds are within 80 kpc of
NGC~3263, then they should be ionized by NGC~3263's UV radiation
field.  Thus we expect that \cld's southeast diagonal component should
be observable using the Fabry-Perot staring method. 

The position of HII galaxy WPV060 co-incides with one of \cld's prominent
density enhancements. However our optical spectra shows that WPV060's
heliocentric systemic velocity 
differs from the central velocity of the density enhancement by 140 \kms. Since
there is no HI in our data at WPV060's systemic velocity, it is difficult to
prove an association.  Thus there is no support for a scenario in which the
enhancement could be a tidal dwarf galaxy.

It is not surprising that bright stars and stellar systems (such as dwarf
galaxies and globular clusters) are not evident in \cld. Both its probable
origin and the current conditions are not conducive to star formation.  A likely
origin is suggested by the fact that the galaxies in the group are tidally
disturbed and that \clds is reminescent of debris in the Virgo cluster.  Thus
\clds could have resulted from the tidal field of the NGC~3256 Group stripping
tidal tails from a pair of interacting galaxies.  In this case the distribution
of stars in the debris could be extremely attenuated or the tidal debris could
have been pulled exclusively from the outer disks' HI gas.  With respect to
current conditions, our exploration indicates that stars will not currently form
out of this debris. Although the enhancements within \clds contain sufficient
mass ($\rm \sim10^{8}\mo$) to be considered possible progenitors of globular
clusters or dwarf galaxies, their internal pressures (P/k $\rm \leq 100\ K\
cm^{-3}$) are less than the $\rm 10^{6 - 8}\ K\ cm^{-3}$ required to instigate
the formation of bound stellar structures.  Also we used the Bonnor-Ebert
formulation to derive a theoretical mass that the enhancements would require in
order to form stars.  The mass observed in HI emission is less than this
required mass, by factors ranging from 6-44.


If an event does not occur to compress \clds then we speculate that the density
enhancements may evolve into the type of object which could be identified with
HVCs. Our investigation shows that the surrounding diffuse material would need
to be stripped away from the density enhancements, say by the overall tidal
field of the NGC~3256 Group of galaxies, and that the enhancements themselves
would need to become more diffuse. The HVC identification would be more likely
if \clds is in the vicinity of NGC~3263, in which case its constituents may
suffer a fate similar to that proposed for the Magellanic Stream. That is, it
may dissolve in NGC~3263's hot gas halo. Alternatively, if \clds is sufficiently
far from an individual galaxy, the 
naked enhancements may fall towards the centre-of-mass of NGC~3256 Group. These
tidal remnants would then mimick the accretion of primordial building blocks
into the potential well of the group's dark matter halo.

\bigskip

Discussions with Jason Fiege were exceedingly useful
and with Dean McLauglin were inspiring. We thank Marija Vlajic for her
contribution to Fig.~\ref{uvfld}.  JE acknowledges
the support from Australia Telescope National Facility in the form of
a Distinguished Visitor Award and from the Canadian National Science
and Engineering Research Council. JBH acknowledges a Federation
Fellowship from the Australian Research Council.

\begin{figure}
\epsscale{.8}
\plotone{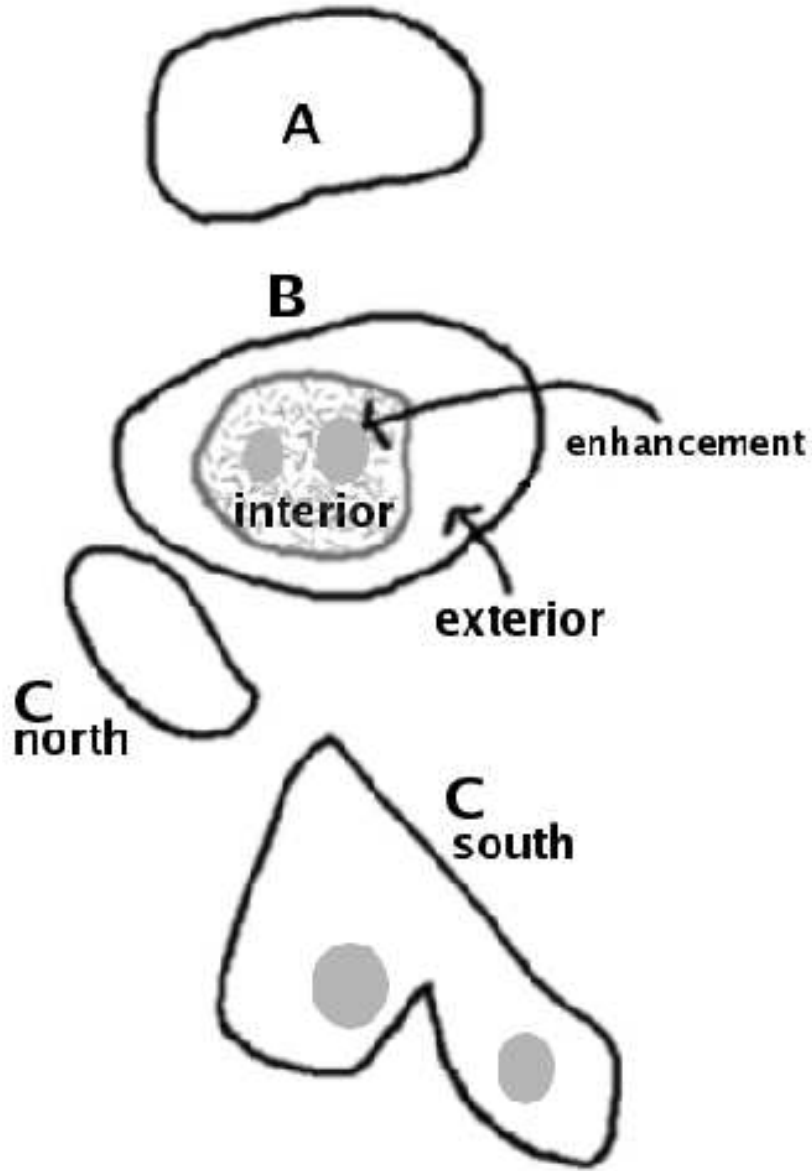}
\epsscale{1}
\figcaption[f2.eps]{Schematic of the components of
\clds as derived from the HI cube (Table~\ref{atcalog}).  Diffuse cloud
components are labelled A, B and C and are apparent in 
Fig.~\ref{zoomcloud}. Note that B appears to have an
exterior diffuse envelope surrounding a more structured interior
component (enclosed by the grey contour); the pressures of the
envelope and the structured interior are compared in
\S~\ref{discusspressure}.  The structured interior in turn contains 2
enhancements.  C has a northern diffuse section and a southern section
which contains 2 enhancements. The models of the enhancements,
described in \S~\ref{obsenhancements}, are represented here using
grey ellipses.
\label{schematic}}
\end{figure}

\begin{figure}
\epsscale{.8}
\plotone{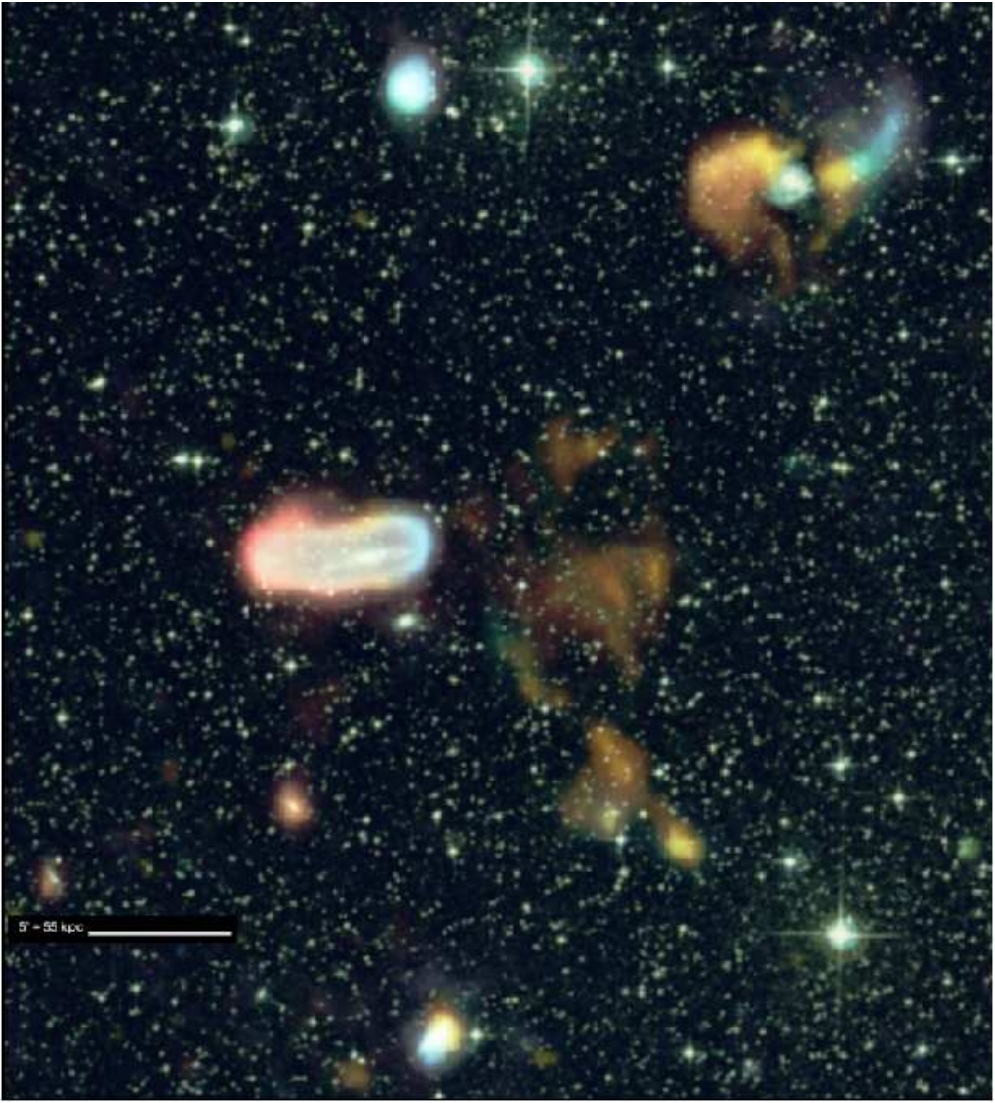}
\epsscale{1}
\figcaption[f3.eps]{HI velocity field in the
NGC~3256 Group of galaxies.  The HI cubes 
were the same as those used
for creating the moment maps in Fig.\ref{HIoverlay}. 
For each of these cubes individual
integrated intensity maps were created for 8 separate velocity ranges
between 2621 to 3337 \kms.
Initially each map was displayed
with a linear intensity stretch and adjusted in order to emphasize the
brighter features in the cloud, generating a greyscale image. Each
greyscale image was subsequently colour coded such that blue
represents blueshifted 21cm emission and red represents redshifted
emission. The radio emission maps from both cubes have been overlaid,
using the screen algorithm \citep{rlfek04}, on colourized Digitized
Sky Survey images from the blue and the red wavelength regimes. 
(Reprinted, with permission, from the cover of I.A.U. Sym 217
2004 ``Recycling Intergalactic \& Interstellar Matter'', editors
Pierre-Alain Duc, Jonathan Braine, and Elias Brinks, A.S.P.~217.)
\label{velfldoverlay}}
\end{figure}

\begin{figure}
\plotone{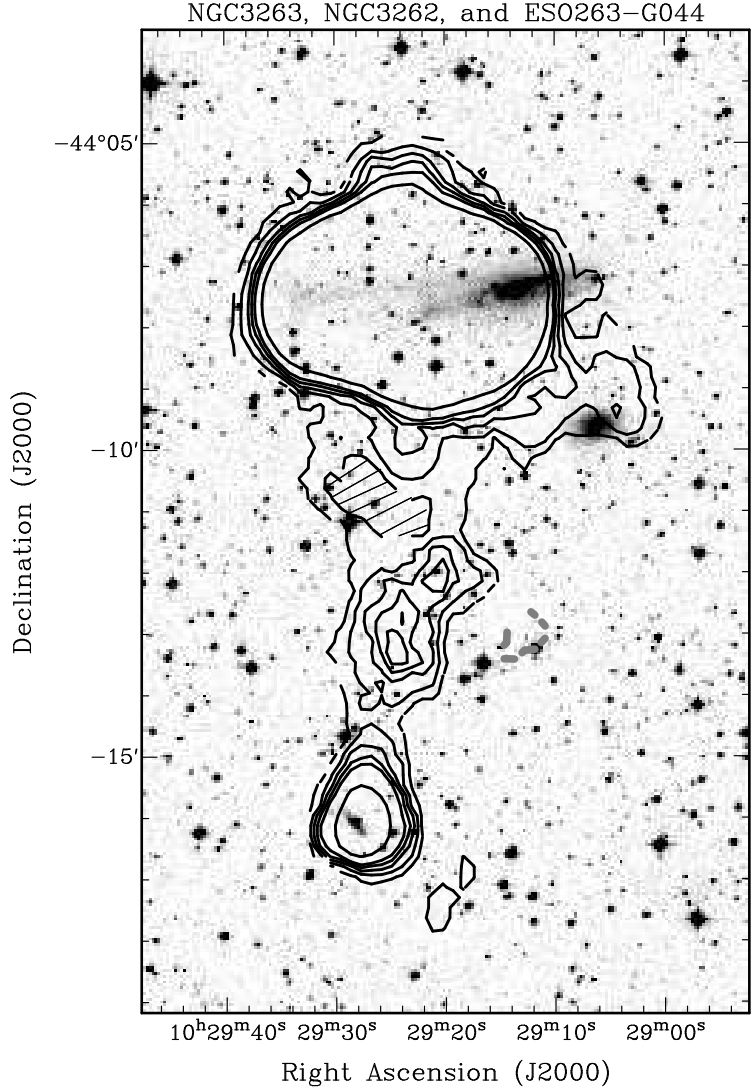}
\figcaption[f4.eps]{The possible HI bridge
between NGC~3263 and ESO~263-G044. Contours from an HI intensity map, 
integrated over the limited velocity range of 
3025 to 3171 \kms,  overlay a greyscale Digitized Sky Survey image.
(Values between $\pm$ 
3 mJy  beam$^{-1}$ (that is, $\pm$ 3 $\sigma$) were clipped from the zeroth 
moment map.)  
West (right) of NGC~3263 lies NGC~3262 while ESO~263-G044 lies to
the south (bottom).  
The HI intensity contours delineating the bridge and outlining the 
galaxies are  60 120  180 240 300 480 mJy $\rm beam^{-1}\ \times$ \kms.
The dashed grey contour (to the right of the bridge)
is -30 mJy $\rm beam^{-1}\ \times$ \kms. No HI emission
resides in the lightly hashed area. See \S~\ref{comparediffuse} for a more
detailed description. 
\label{bridge}}
\end{figure}

\begin{figure}
\epsscale{.8}
\plotone{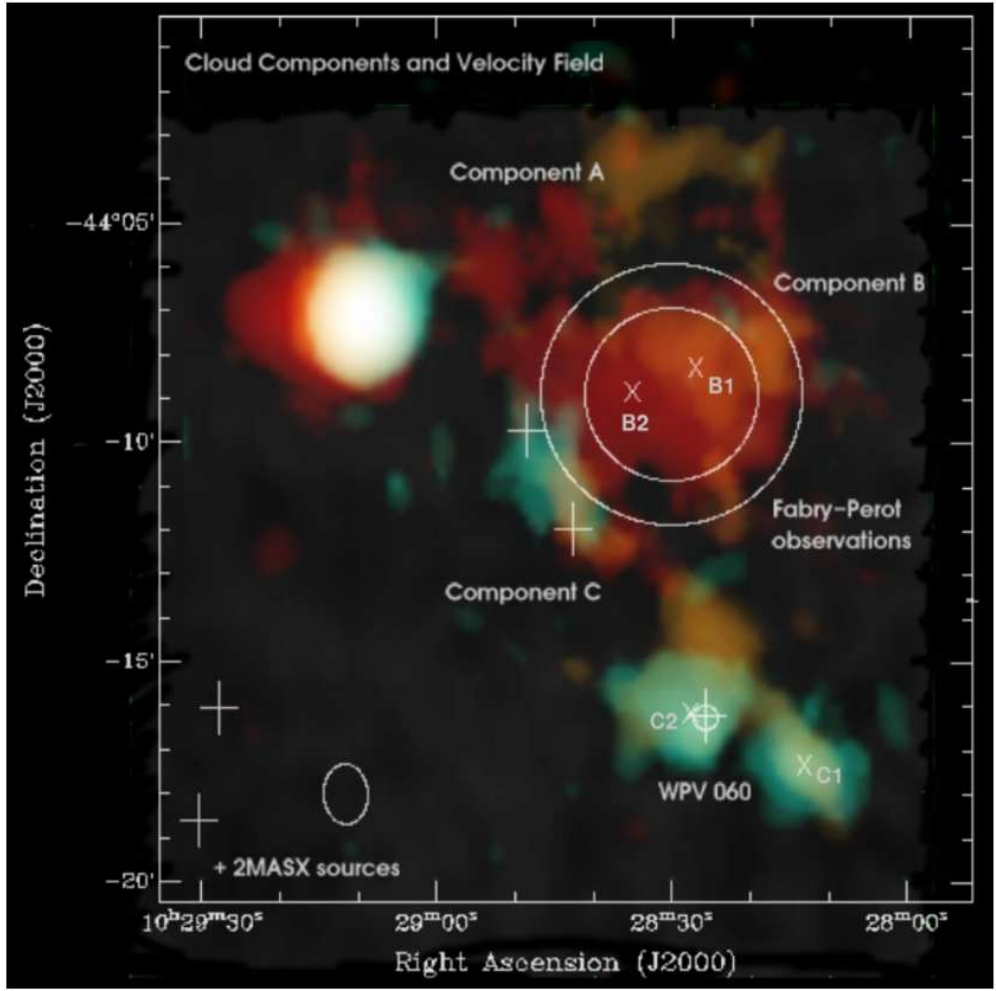}
\epsscale{1}
\figcaption[f5.eps] {Features in the Mean HI Velocity Field
of \cld.  Three integrated intensity maps were made from a lower
velocity resolution (10 \kms) HI cube than the cube presented in 
Table~\ref{atcalog}. The colour
turquoise was assigned to the map covering velocities from
roughly 2773 through
2830 \kms; gold to 2831 - 2858 \kms; and dark red to 2859 -
2940 \kms. 
The bright white feature to the east is part of NGC~3263.
Component A arcs horizontally from NGC~3263 to the west; component B
is the elliptically shaped emission cloud below A; and component C runs on a
diagonal from the east edge of B to the southwest.  These 3
morphological components are described in more detail in
\S~\ref{detectcloud} and sketched in Fig.~\ref{schematic}.  
X-shaped crosses, with letter designations, mark
positions of substructures found in the higher velocity resolution
spectra of the cube listed in Table~\ref{atcalog}
and described in \S~\ref{resultsenhance}.  Plus-shaped
crosses mark 2MASS infra-red sources.  WPV060, designated by
a  circle, is a starburst galaxy
also discussed in \S~\ref{resultsenhance}.  For scale, ten seconds 
of right ascension,
using a declination of -44\degr09\arcmin, is about 1.8\arcmin\ and, at a
distance of 37.6 Mpc, 1\arcmin\ equals 11 kpc.\label{zoomcloud}}
\end{figure}

\begin{figure}
\plotone{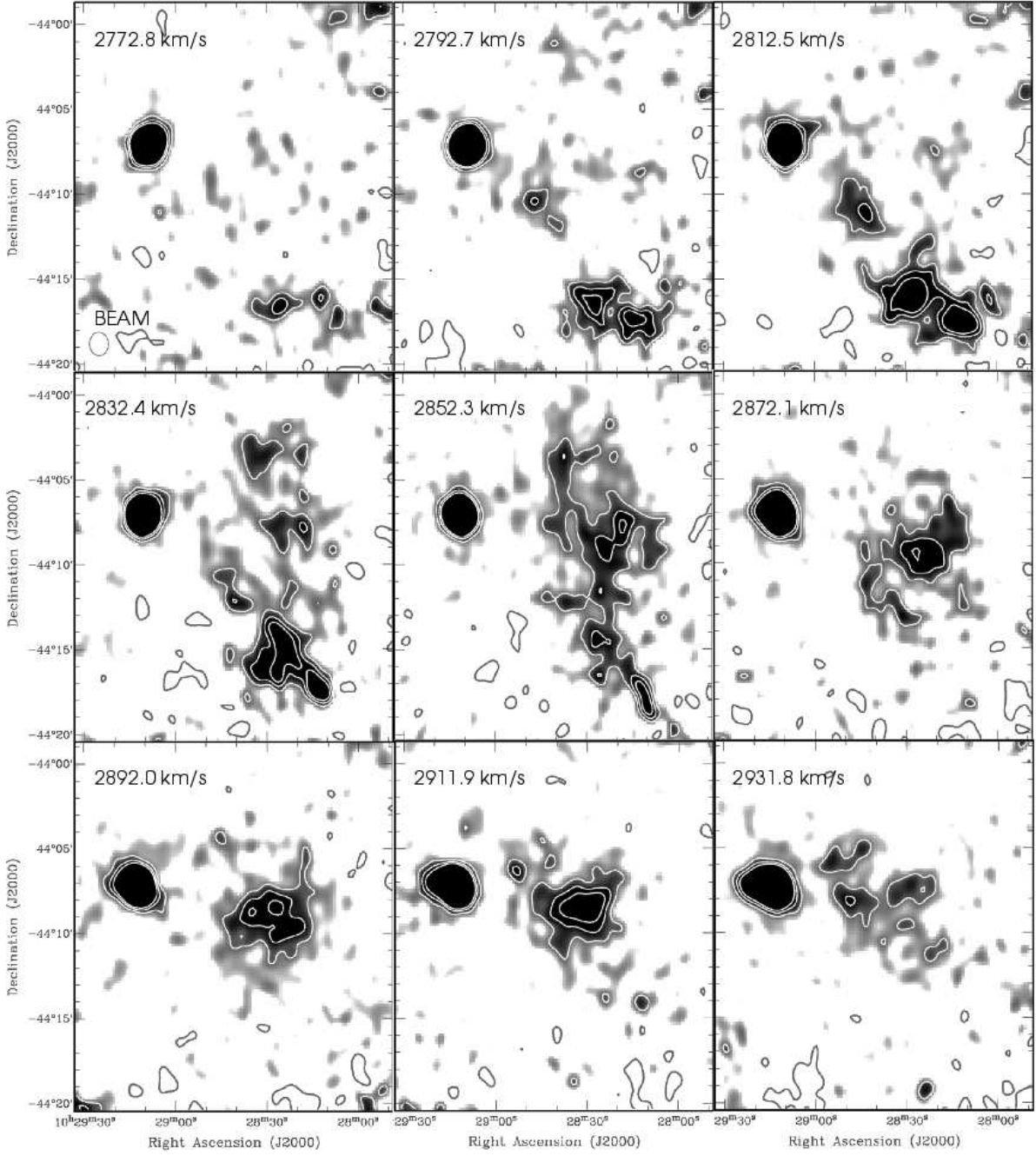}
\epsscale{1.0}
\figcaption[f6.eps]{HI channel maps of \cld.
Each plane consists of three combined channels from the 
HI cube (Table~\ref{atcalog})
and has a velocity width of $\sim$20 \kms.
The greyscale ranges from 1 mJy $\rm beam^{-1}$ ($\sim 1 \sigma$) 
to 5 mJy $\rm beam^{-1}$.
The dark contours are -3 mJy $\rm beam^{-1}$ while the white contours
are 3, 5, and 7 mJy $\rm beam^{-1}$. The enhancements in component C are clear
at 2792.7 \kms\  while the enhancements in component B
are evident at 2892 \kms. Note that all 3 components appear associated
at 2852.3 \kms. The strong circular feature to the east (centered near
10\hr29\mn10\fs, -44\degr07\arcmin) that appears in every panel is
part of NGC~3263. \label{velfldmosaic}}
\end{figure}

\begin{figure}
\plotone{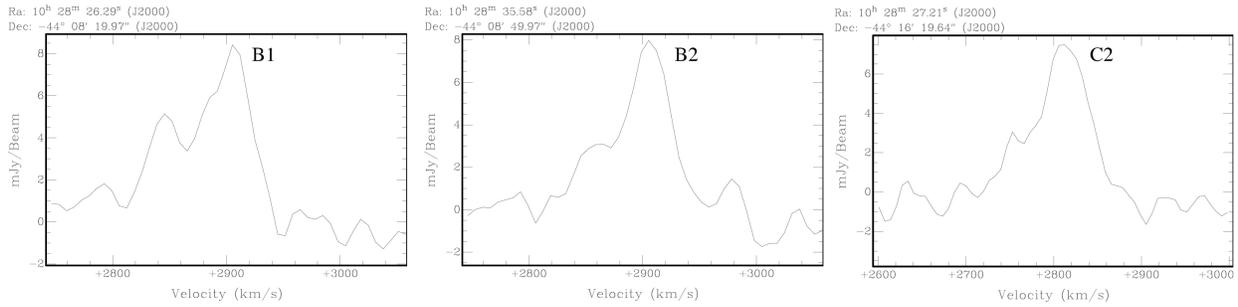}
\figcaption[f7.eps]{Velocity 
profiles of the enhancements.  These profiles, while covering the spatial 
region of the enhancements, include emission in the velocity range of the 
host component (\S~\ref{obsenhancements}/\ref{resultsenhance}). 
Each labelled peak is associated with an enhancement's velocity range assessed 
visually within the data cube. \label{enhanceprof}}
\notetoeditor{Please rotate this figure to landscape mode so that the lines show
well. Also I could not get it to be on a separate page.}\bf{Editor: Please rotate this figure to landscape mode so that the lines show
well. Also ensure that it is on a separate page.}
\end{figure}

\begin{figure}
\plottwo{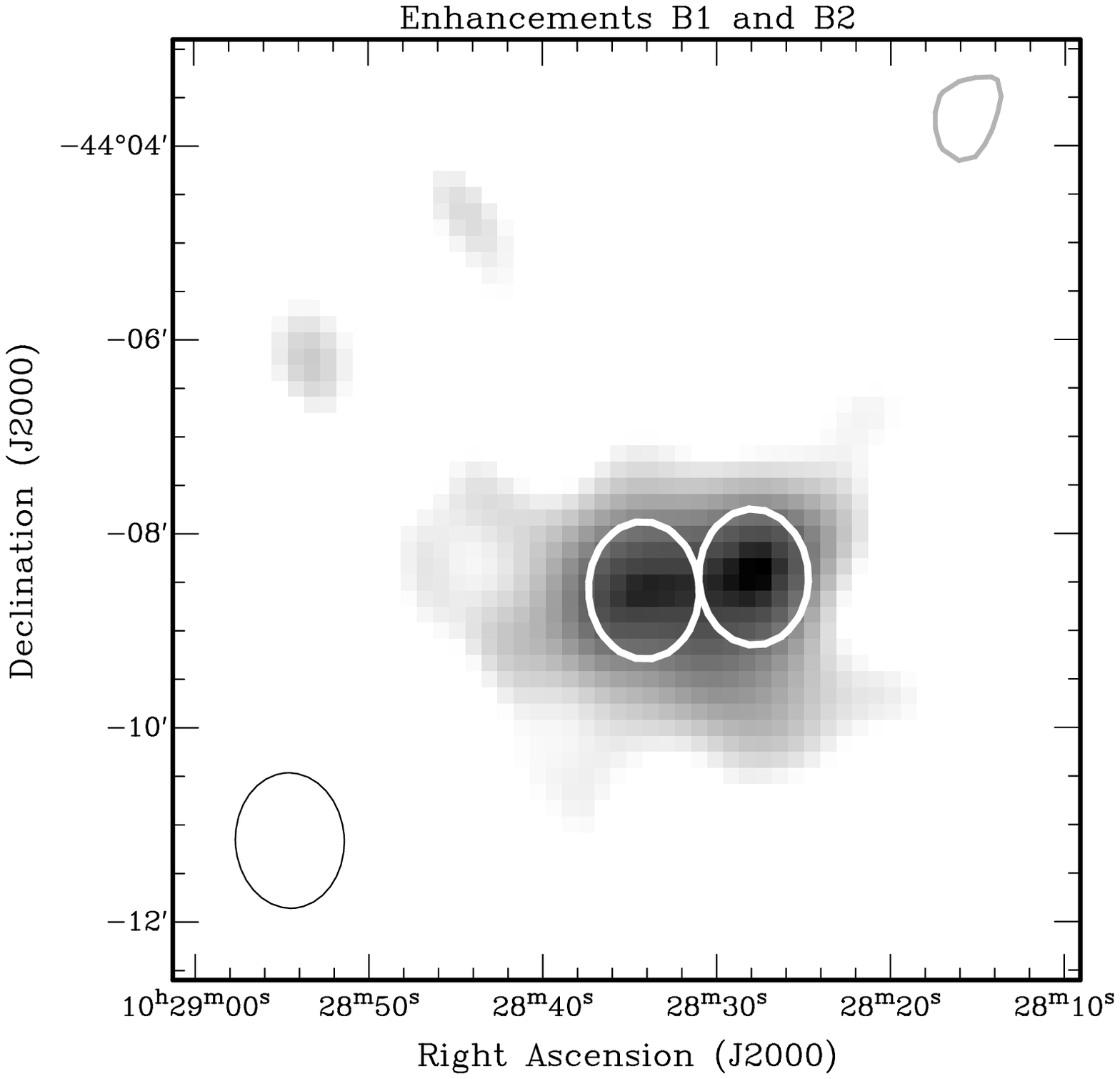}{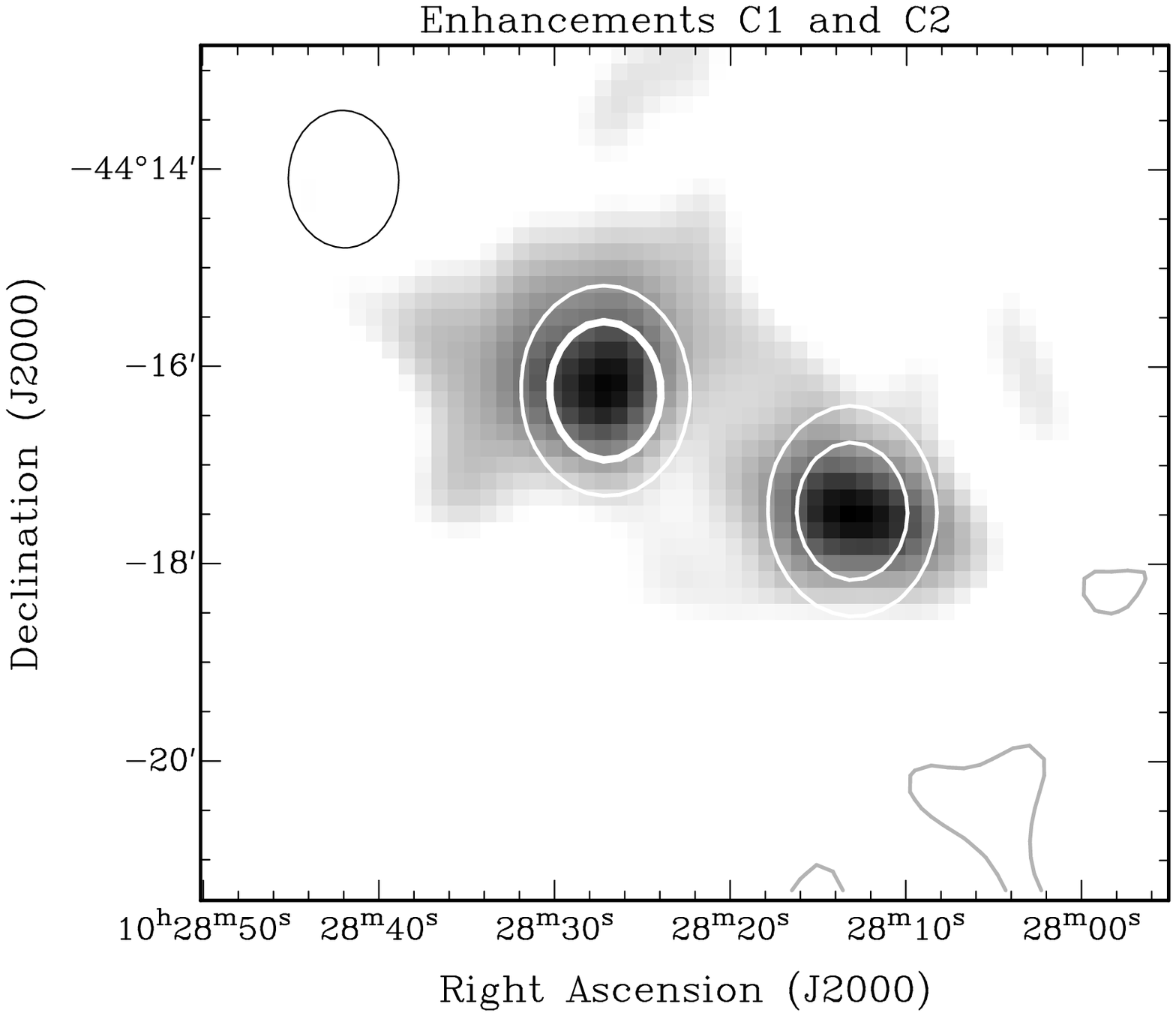}
\figcaption[f8ab.eps]{Integrated HI intensity maps of the
enhancements.  The candidate enhancements in each cloud 
appear in integrated intensity maps constructed from cubes 
in which the flux level of the surrounding diffuse cloud has been 
subtracted 
(\S~\ref{obsenhancements}). For the left-hand image containing B1 and B2,
the non-linear greyscale displays values in the range of 
0.3 to 200.1 mJy $\rm beam^{-1}\ \times$ \kms. 
For the greyscale image containing C1 and
C2 the values are  0.6 
to 246.7 mJy $\rm beam^{-1}\ \times$ \kms. For 
comparison of noise features
with the shapes of the enhancements, we show  grey contours 
which trace -190 mJy $\rm beam^{-1}\ \times$ \kms.
The white contours outline the 50\%
contour of the Gaussian model fit used to measure the flux in each
enhancement in Cloud B and the 20\% and 50\% levels in Cloud C; see
\S~\ref{obsenhancements}. 
The black ellipse, in the lower left corner,
outlines the FWHM of the synthesized
beam.
\label{plotclumps}}
\end{figure}

\begin{figure}
\plottwo{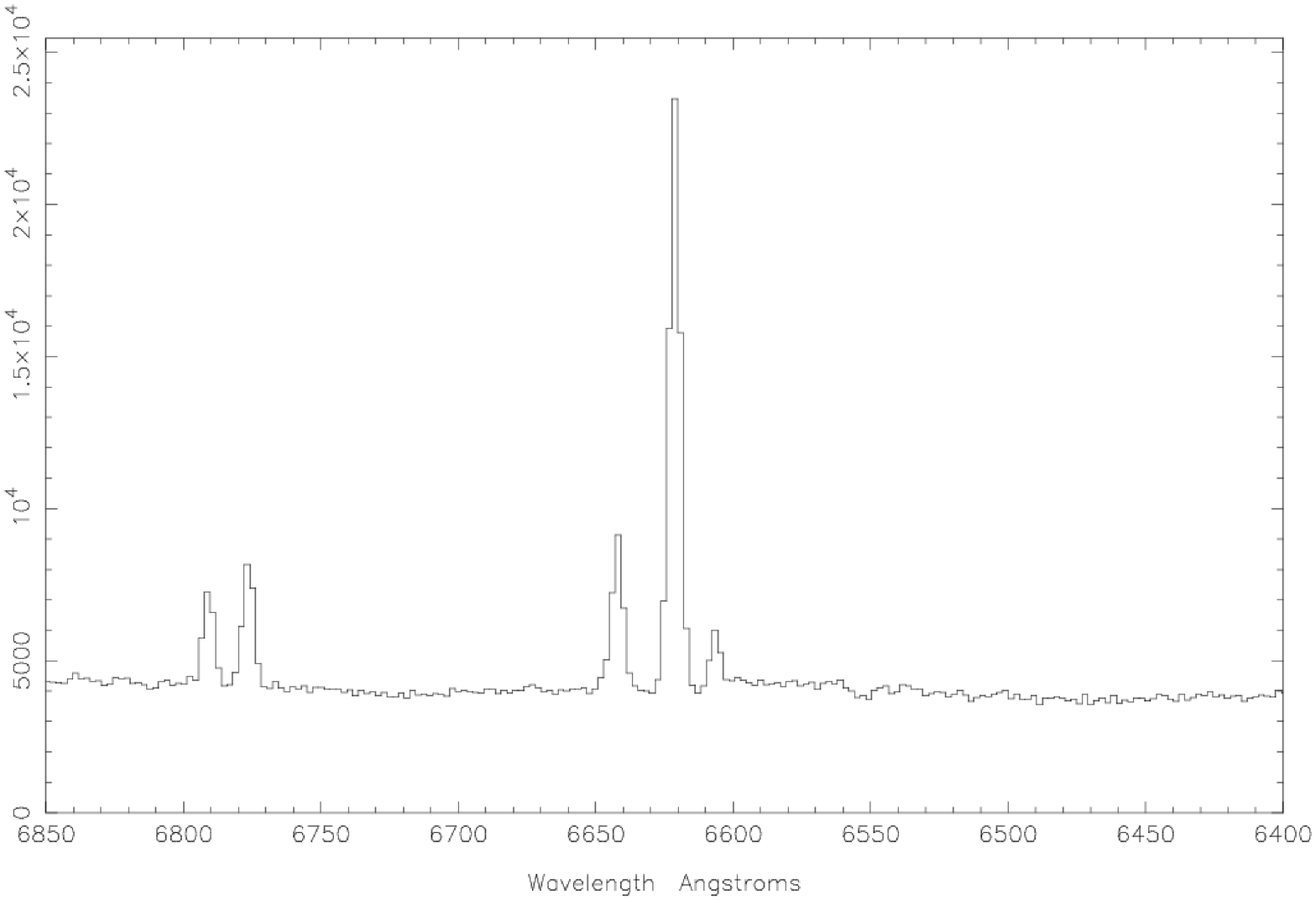}{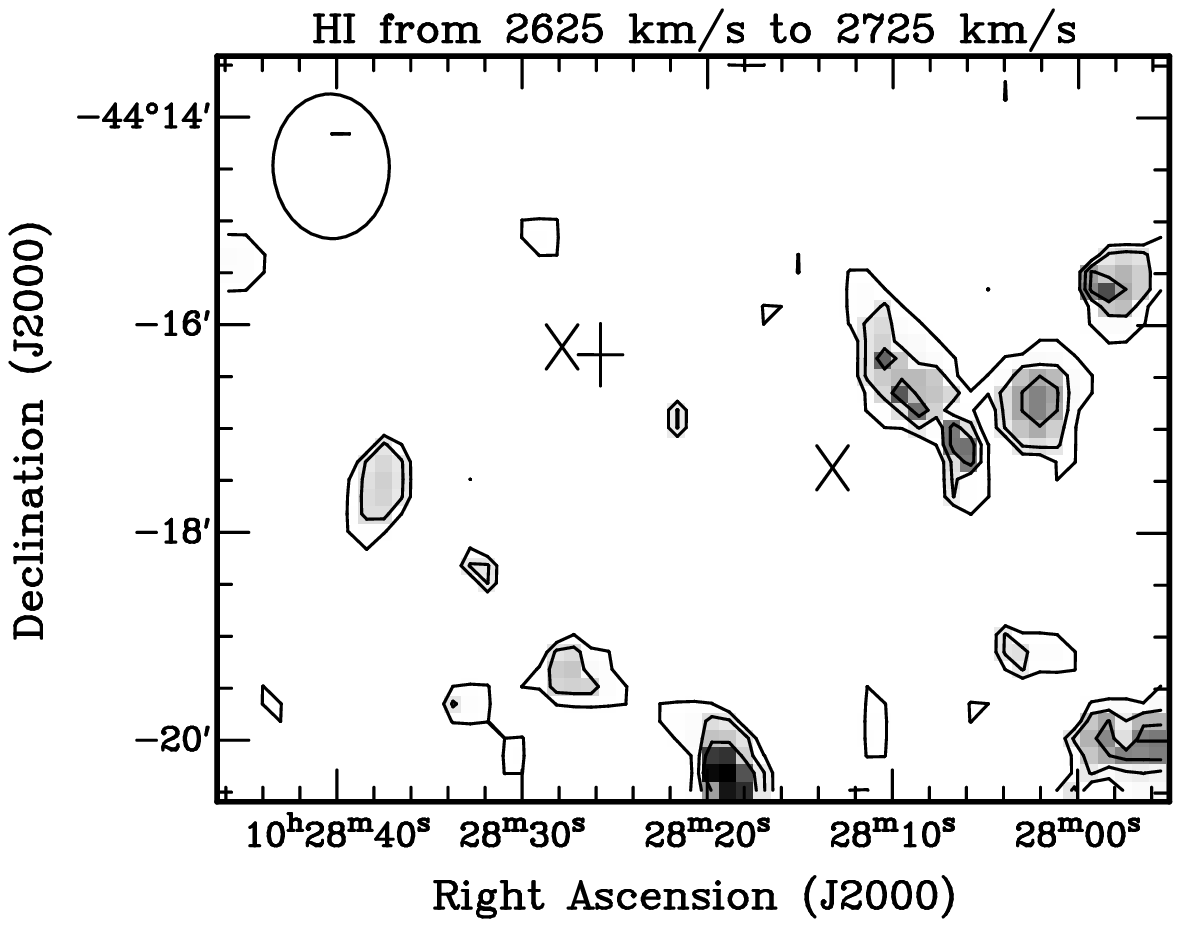}
\figcaption[f9ab.eps]
{Galaxy WPV060.  See \S~\ref{resultsenhance} for discussion. 
Details about the optical observations for the spectrum on the left
are described in \S~\protect\ref{opticalobs}. See \S~\ref{resultsenhance} 
for the discussion associated with this figure. The redshifted optical emission
lines give a heliocentric systemic velocity for WPV060 of 2675 $\pm$ 6
\kms.  
The map on the right integrates the HI intensities in the velocity range
between 2625 and 2725 \kms, which is centred on the peak optical 
velocity of WPV060. This demonstrates that, at the spatial scale of this
HI cube (Table~\ref{atcalog}), 
there is no
emission associated with this HII galaxy; the features in the plot are noise.
The beam is represented
by the ellipse in the upper left corner. Only intensities above 
3 $\sigma$ where used; the contours are 4 (= 3 $\sigma$), 
8, and 12 mJy beam$^{-1}$. The position of C2 is marked by
the $\times$ on the left and C1 is the $\times$ on the right. The 
+ marking the position of WPV060 is larger than the galaxy's 
diameter of $\sim$7\arcsec. 
\notetoeditor{please rotate this so the figures are side-by-side in
landscape orientation.}
\label{haspec}}

\notetoeditor{Please rotate this figure to landscape mode so that the lines show
  well. Also I could not get it to be on a separate page.}\bf{Editor: Please rotate this
  figure to landscape mode so that the lines show well.}

\end{figure}

\begin{figure}
\plotone{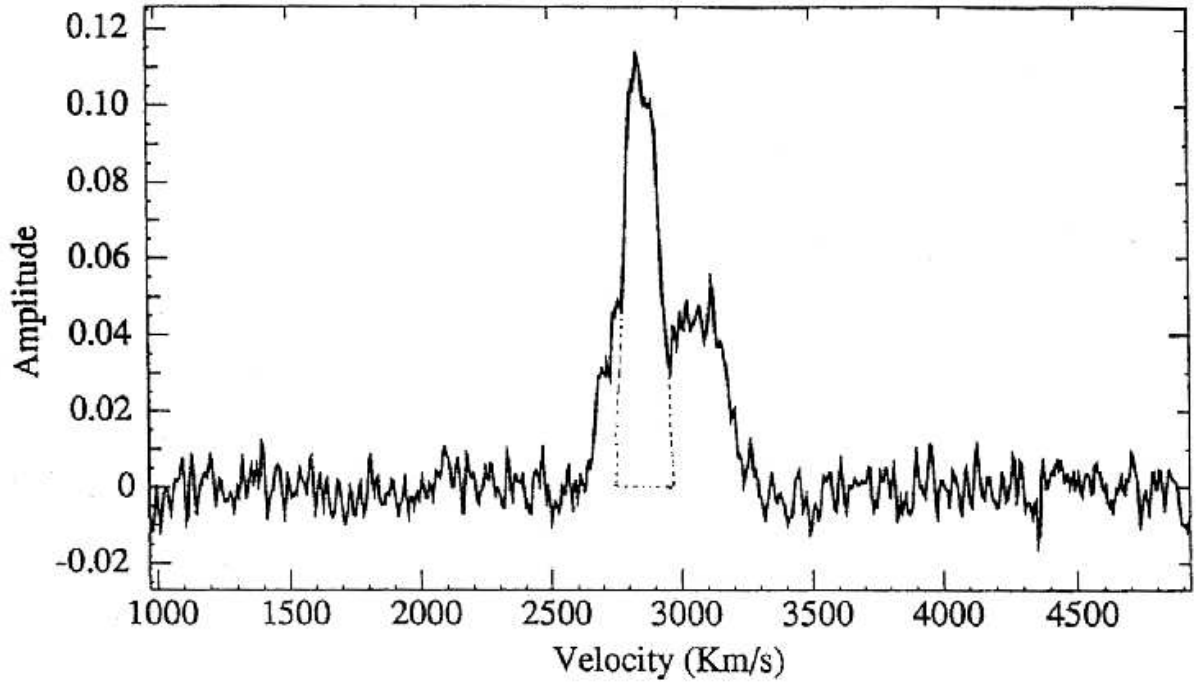}
\figcaption[f10.eps]{Parkes Single-Dish Telescope Profile. Amplitude is 
in units of Jy. 
The on-target position is
${\alpha}$~=~10\hr 28\mn 27\fs7, ${\delta}$~=~-44\degr 09\arcmin 26\arcsec 
(J2000). The area under the peak,
enclosed in a dashed line,
is believed to be associated with \clds rather than NGC~3263; more 
detail is given in \S~\protect\ref{derived} and \citealt{eng94}. This area
is used to determine the mass of \clds in \S~\protect\ref{derived}. 
\label{parkesprofile}}
\end{figure}

\begin{figure}
\plotone{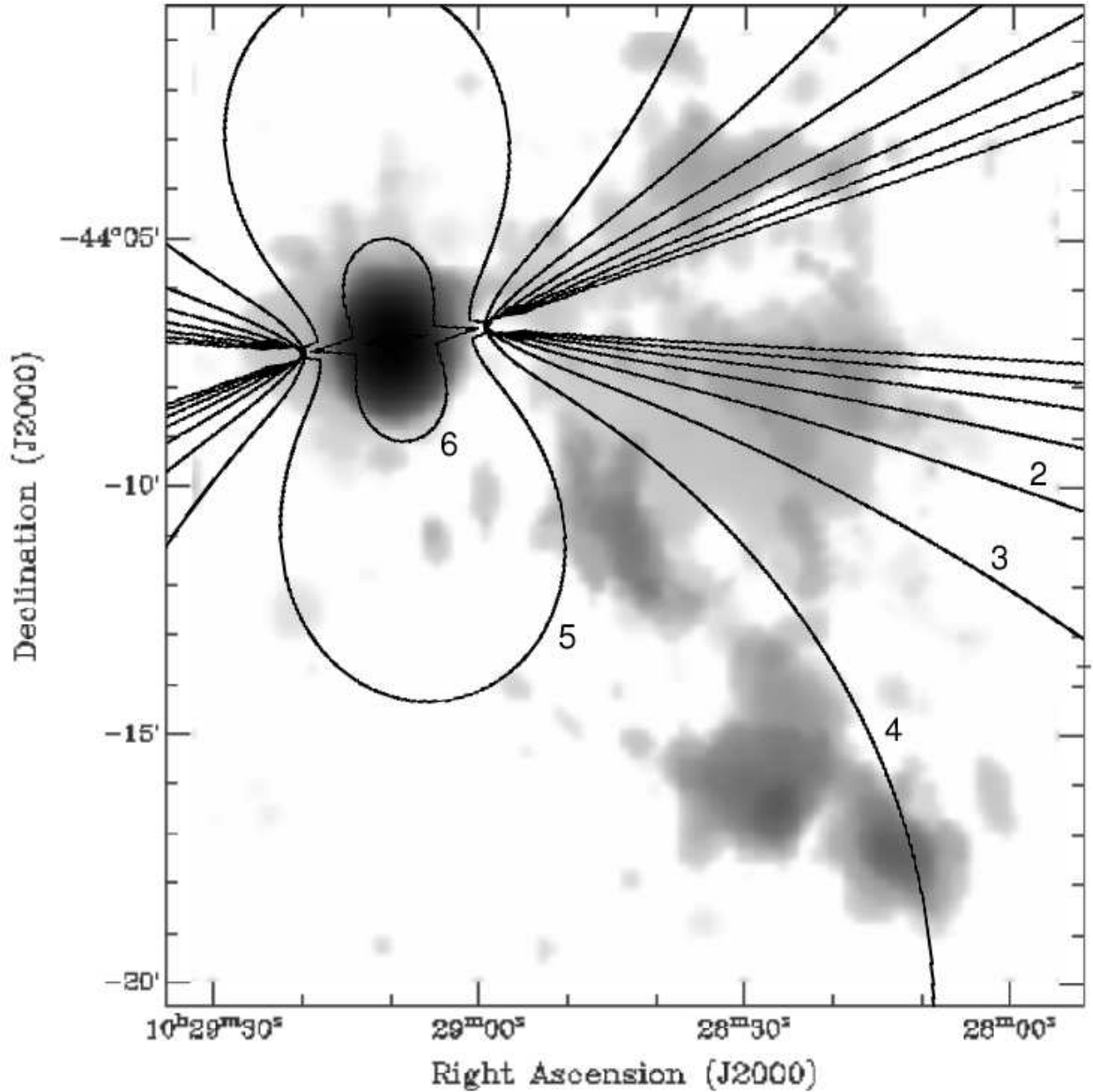}
\figcaption[f11.eps]{The distribution of the ionizing
radiation above and below the plane of NGC~3263
(position angle = 10 degrees) superimposed on a B\&W rendition of
the HI velocity data in 
Fig.~\ref{zoomcloud}. The model is overlaid on the optical 
centre of NGC~3263 and the lines show a
cross-section through the radiation field. The contours
represent lines of equal ionizing flux. The units are
log(photons cm$^{-2}$ s$^{-1}$). See \S~\protect\ref{uvmodel}.
\label{uvfld}}
\end{figure}


\clearpage
\begin{deluxetable}{lcccc} 
\tablecolumns{2} 
\tablewidth{0pc} 
\tablecaption{Australia Telescope Compact Array HI Data Cube: 
Characteristics 
of the position-velocity cube created from 21-cm spectral line data.  
Subcubes with smaller velocity ranges were produced for 
the various analyses described in \S~\ref{21measure}. 
\label{atcalog} } 
\tablehead{
\colhead{ Parameter} 
& \colhead{HI Cube} 
 }
\startdata 
{ Synthesized Beam} 
  (arcsec)     &  84 x 67 \\ 
{ Beam Position Angle}& 2.5$^{\circ}$ \\ 
{ Channel Width }
   (km $\rm s^{-1}$) &  6.6   \\ 
{ Weighting} &  natural \\
{ Measured $\rm r.m.s.$}\tablenotemark{a} 
(mJy $\rm beam^{-1}$) &  1.3  \\
{ Total Velocity Range (km $\rm s^{-1}$)} & 2330 to 3416
\enddata 
\tablenotetext{a}{The $\rm r.m.s.$ was determined before correction  
for the primary beam.}
\end{deluxetable}

\begin{deluxetable}{lccc}
\tablecolumns{4} 
\tablewidth{0pc} 
\tablecaption{AAO Fabry-Perot Integration Times. These observations
attempt to detect \ha in Cloud B. See \S~\ref{opticalobs} for 
details such as the coordinates of the field centre, 
central wavelength and passband width. \label{aaocatalog}}
\tablehead{
\colhead{Target}
& \colhead{Exposure Time (sec)}
& \colhead{Airmass}
& \colhead{Description}}
\startdata 
\Clds &  1200 & 1.3 & target field\\
\Clds &  1200 & 1.1 & target field\\
\Clds &  1200 & 1.1 & target field\\
\Clds &  1200 & 1.0 & target field\\
EG 193 & 120 $\rm \times$ 6& 	1.5& photometric standard\\
\enddata 
\end{deluxetable} 

\begin{deluxetable}{lcccccc} 
\tablecolumns{6} 
\tablewidth{0pc} 
\tablecaption{Measured and derived characteristics of the 3 diffuse 
components of \cld.
Measurements, analysis, and assessment of uncertainties of the observables
are in \S\ref{21measuremom}. 
Derivations of mass, pressure and uncertainties are described in
\S\ref{analysisderive}.
\label{combinedtable}}
\tablehead{
\colhead{}
&\multicolumn{5}{c}{Cloud Component}
\\
\colhead{}
&\colhead{A}
&\multicolumn{3}{c}{B}
&\colhead{C}
\\
\colhead{}
&\colhead{}
&\colhead{Whole}
&\colhead{Interior\tablenotemark{a}}
&\colhead{Exterior}
&\colhead{}
\\}
\startdata 
{Heliocentric Systemic}&  
2842$\pm$ 4& 2885 $\pm$ 6 & 2905 $\pm$  6 &\nodata & 2813 $\pm$ $>$7\\
 {\ Velocities (\kms)} &&&&&\\
{FWHM (\kms)\tablenotemark{b}}& 41 & 97 & 51&97&\nodata \\
{Velocity Range}& 2832-2852 &2832-2938&  2872-2932&\nodata  &2786-2839\\
{\ (\kms)} &&&&&\\
{Length (arcsec)\tablenotemark{c}} &220 $\pm$ 26& 368 $\pm$ 18&175 $\pm$ 5 &
\nodata & 760 $\pm$ 10\\
{Width (arcsec)}&140 $\pm$ 14&  300 $\pm$ 20  & 130 $\pm$ 9  &\nodata & 260 $\pm$ 10\\
{Flux Density}& 0.47 &  4.2&  0.77&\nodata  & 4.9 \\
{\ (Jy \kms)\tablenotemark{d}}&&&&&\\
{Column Density}& 1.10 $\times  10^{19}$&6.83  $\times  10^{19}$
    & 7.45 $\times  10^{19}$&\nodata  &6.67 $\times  10^{19}$ \tablenotemark{e}\\
{\ (atoms cm$^{-2}$)\tablenotemark{d}}&&&&&\\
{Peak Column Density}& 2.62  $\times  10^{19}$& 1.21 $\times  10^{20}$ &
&\nodata & 1.70$\times  10^{20}$\\
{\ (atoms cm$^{-2}$)\tablenotemark{d}}&&&&&\\
{Mass ($\mo$)\tablenotemark{d}} & 1.6 $\times 10^{8}$ &14.0 $\times 10^{8}$
&2.5 $\times 10^{8}$& 11.5 $\times 10^{8}$& 16.3 $\times 10^{8}$\\
{Pressure/k ($\rm K\ cm^{-3}$)\tablenotemark{f}}&5 & 100 & 50 &91& 39\\
\enddata 
\tablenotetext{a}{\footnotesize The central region of component B. See
\S~\ref{analysisderive} and Fig.~\ref{schematic}.  }
\tablenotetext{b}{\footnotesize The uncertainty (\S~\ref{21measuremom}) is 
10 to 15\%. With respect to Cloud C see the end of \S~\ref{detectradio}.}
\tablenotetext{c}{\footnotesize At a distance of 37.6 Mpc, 
1$^{\rm \prime \prime}$ is equivalent to 182 pc.} 
\tablenotetext{d}{\footnotesize The uncertainties are(\S~\ref{21measuremom}) :  
flux density $\leq 30\%$;  
column density $\leq 15\%$; peak column density $\sim 40\%$.
The uncertainty for mass is $\sim 33\%$ (\S~\ref{detectradio}).}
\tablenotetext{e}{\footnotesize For cloud C the column density value is the
average of the north and south components which, separately, are $\rm
4.58\ \times\ 10^{19}\ atoms\ cm^{-2}$ (north) and $\rm
8.76\ \times\ 10^{19}\ atoms\ cm^{-2}$ (south).} 
\tablenotetext{f}{\footnotesize  The uncertainty is 43\%
(\S~\ref{analysisderive}). Pressure has been divided by the Boltzmann Constant k.} 
\end{deluxetable}

\begin{deluxetable}{p{1.25in}ccccc} 
\tablecolumns{6} 
\tablewidth{0pc} 
\tablecaption{Measured HI characteristics of the enhancements in \cld.
See Figure~\ref{zoomcloud} for the relative positions of these
enhancements.  Measurements, 
analysis, and assessment of uncertainties 
are described in \S\ref{obsenhancements}\label{tabclumpobs}.} 
\tablehead{
\colhead{ Enhancement} 
& \colhead{B1}
& \colhead{B2}
& \colhead{C1z}
& \colhead{C1y}
& \colhead{C2}\\
 }
\startdata 
Position{\footnotesize(J2000)}\tablenotemark{a}   &  &  &  &    \\
  $\alpha$ {\footnotesize(\hr \mn \st)} & 10 28 27.2 &  10 28 35.3 
          & 10 28 13.2 & 10 28 13.2 & 10 28 27.8  \\ 
 $\delta$ {\footnotesize(\degr \arcmin \arcsec)} & -44 08 17 &  -44 08 50
          & -44 17 28 & -44 17 15 &  -44 16 13  \\
  &  &  &  &    \\
{Central~Velocity\tablenotemark{b}  \break {\footnotesize(\kms)}} &
           2906 $\pm$ 3 & 2908 $^{+3}_{-9}$ 
              & 2807$\pm$7 & 2835$\pm$7 &2815 $\pm$ 3\\
  &  &  &  &    \\
Peak Intensity r.m.s.\tablenotemark{c}  & 3 & 2 &4 & 3  & 3  \\
  &  &  &  &    \\
FWHM\tablenotemark{d} \ \  {\footnotesize(\kms)}
  & 31 $\pm$ 5 & 28 $\pm$ 2 
  & 20 $\pm$ 6 
  & 21 $\pm$ 6       &  44 $\pm$ 1 \\ 
  &  &  &  &    \\
{Integrated\nobreak  ~Flux Density\tablenotemark{e} 
{\footnotesize(Jy$\times$\kms)}}&
0.26 &0.23 &0.28&0.22&0.40\\ 
\enddata 
\tablenotetext{a}{\footnotesize 
The uncertainty in position is 26\arcsec (\S~\ref{obsenhancements}).
}
\tablenotetext{b}{\footnotesize 
These heliocentric radial velocities are the central velocities 
of single Gaussian profiles, except for C1. C1's  two  
velocity peaks are labeled 
z and y and blend at the 50\% level.  Note, 
B2 is visually most delineated at 2897 \kms.
} 
\tablenotetext{c}{\footnotesize  This is the number of r.m.s. at which
the peak intensity occurs above the 
mean intensity of the surrounding diffuse host component.
See \S~\ref{obsenhancements}.}
\tablenotetext{d}{\footnotesize The uncertainty includes differences
  between using smoothed and unsmoothed profiles.}
\tablenotetext{e}{\footnotesize  The uncertainty is $<$
 0.06 Jy$\times$\kms; \S~\ref{obsenhancements}.}
\end{deluxetable} 

\begin{deluxetable}{lccc} 
\tablecolumns{4} 
\tablewidth{0pc} 
\tablecaption{Derived Parameters for Enhancements.
The integrated flux density used 
to derive the mass associated with HI emission is listed 
in Table~\ref{tabclumpobs}. 
The Bonnor-Ebert mass, $\cal M\rm(r_c)$, associated with
an isothermal sphere on the verge of collapse, 
is derived in \S~\ref{discussmass}. \label{tabclumpsderive} } 
\tablehead{
\colhead{ } 
& \colhead{}
& \colhead{Enhancements}
& \colhead{}\\
\colhead{ Parameter} 
& \colhead{B1}
& \colhead{B2}
& \colhead{C2}\\
 }
\startdata 
$\sigma_o$\tablenotemark{a} {\footnotesize(\kms)} &
      13.2     & 11.9 & 18.7  \\ 
{$\mh$\tablenotemark{b} {\footnotesize($\mo$)}}&
0.9$\times 10^8$  &  0.8$\times 10^8$ & 1.3 $\times 10^8$    \\

{$\cal M\rm(r_c)$\tablenotemark{c} {\footnotesize($\mo$)}}&
1.02 $\times 10^9$ & 0.69 $\times 10^9$ & 5.68 $\times 10^9$    \\
$\frac{\cal M\rm(r_c)}{\mh}$ & 11 & 9 & 44 \\
\enddata 

\tablenotetext{a}{ The velocity dispersion in an 
enhancement $\sigma_o$ is derived using the 
FWHM listed in Table~\ref{tabclumpobs}.}
\tablenotetext{b}{The uncertainty is $\rm \leq 29\%$. The measurement is described
in \S~\ref{obsenhancements} and the uncertainty in \S~\ref{resultsenhance}.}
\tablenotetext{c}{The uncertainty is $\rm \leq 43\%$;\S~\ref{discussmass}.}
\end{deluxetable}

%
%

\end{document}